\documentclass{article}
\usepackage{amsmath,graphicx}

\usepackage{amsthm}
\usepackage{amssymb}
\usepackage{xypic}
\usepackage{multirow}
\theoremstyle{plain}
\newtheorem{theorem}{Theorem}
\newtheorem{proposition}[theorem]{Proposition}
\newtheorem{corollary}[theorem]{Corollary}
\newtheorem{lemma}[theorem]{Lemma}

\theoremstyle{definition}
\newtheorem{definition}{Definition}

\newtheorem{remark}{Remark}
\newtheorem{example}{Example}

\def\shf{\mathcal}

\def\col{\mathcal}

\def\argmin{\textrm{argmin}}
\usepackage{hyperref}

\title{Statistical Sheaf Modeling Methods for Measuring Qualia Structures}

\author{Donna A. Dietz \and Michael Robinson}

\begin{document}

\maketitle

\abstract{
  Mathematical \emph{sheaves} are an abstraction that captures the notion that consistency between local properties can lead to a global picture. There are a number of (mathematical) phenomena that sheaves exhibit, which seem tantalizingly close to what one might hope a theory of qualia would explain. Recently, the theory and abstraction of sheaves has been transformed into robust computational tools for understanding complex systems. These tools are now ripe to help qualia researchers forge robust and testable models. Intuitively, as conscious entities, rather than experiencing ``everything all at once,'' our experience is coherent to varying degrees. Sheaves can model this rather nicely. This chapter presents the practical and theoretical considerations of sheaves in data modeling, with a view towards computational models that can support statistical hypothesis testing.
  }

\tableofcontents

\section{Introduction}

Formalizing a problem carefully provides a handhold for modeling,
by identifying which comparisons between variables are admissible.
This chapter shows that we can encode admissible comparisons as mathematical \emph{sheaves},
and then derive practical solutions to problems by building successively more structured consistency checks,
to measure, localize, and interpret the extent of consistency within a model with respect to observations.

A model should be internally coherent before it can be tested against data.
As this chapter proceeds,
we will successively strengthen what we mean by coherence.
We start with assessing structural consistency among model components,
then functional consistency between model state information, and
finally statistical consistency between observations as mediated by the model.
Each step imposes additional constraints, against which we gain inferential leverage.
The weakest consistency checks can be encoded as elementary mathematical structures that are likely familiar to the reader already, such as \emph{directed graphs}.
Stronger consistency checks require increasingly sophisticated mathematical objects, such as \emph{sheaves of pseudometric spaces} and the \emph{assignment distribution}.
Table \ref{tab:outline} summarizes the key definitions that achieve this progression.

When we build models by diagramming the dependencies between variables,
we reveal how complex hypotheses are assembled, and how they must be tested against observational data.
The unifying theme is that \emph{spans}---where several observable variables are predicted from a latent variable---build complex hypotheses.
On the other hand, \emph{cospans}---where distinct variables all predict a common variable---test hypotheses.
Together, spans and cospans describe the architecture of hypothesis construction and testing,
which the chapter expands into a systematic and practical framework for modeling.
Those spans and cospans that are relevant to a problem are captured in its \emph{topology}.
The admissible values and functional relationships between variables are representable by a \emph{sheaf} on this topology.

\begin{table}
  \caption{Tower of consistency checks of increasing inferential power}
  \label{tab:outline}
  \begin{tabular}{|l|l|l|}
    \hline
    Concept & Ref. &Consistency check on...\\
    \hline
    Directed graph & Def. \ref{def:graph}  & Dependency assumptions\\
    Topology & Def. \ref{def:topology} & Locality assumptions \\
    Sheaf & Def. \ref{def:sheaf} & Functional relationships\\
    Section & Def. \ref{def:section} & Agreement between observations \\
    Continuity & Def. \ref{def:continuity} & Quantitative stability of predictions\\
    Consistency radius & Def. \ref{def:consistency_radius} & Observational fit to model\\
    $p$-value & Def. \ref{def:p_value} & Statistical plausibility\\
    \hline
  \end{tabular}
\end{table}

The goal of this chapter is to show how these successive notions of consistency culminate in a practical statistical framework,
in which model-data agreement is quantified by consistency radius and evaluated through hypothesis testing.

Although this chapter includes many small examples to explain the concepts involved,
we will introduce the example of \emph{binocular rivalry} \cite{Cha_2018,Lee_2004,Kovacs_1996}, 
and carry it through the chapter as a sequence of successively richer examples.
These examples follow the tower of modeling constraints outlined in Table \ref{tab:outline}, 
\begin{description}
\item[Directed graph] (Example \ref{eg:binocular_graph}) Shows the relationship between image sense data from the two eyes and the overlap region between them,
\item[Topology] (Example \ref{eg:binocular_topology}) Adds a unified latent scene ``in the brain'' or ``in the world,''
\item[Sheaf] (Examples \ref{eg:binocular_sheaf} and \ref{eg:depth_sheaf}) Adds the response to light from each receptor in the retina and the functional relationships between them,
\item[Section] (Example \ref{eg:binocular_section}) Infers a unified scene from both eyes if they agree about the scene,
\item[Continuity] (Example \ref{eg:binocular_continuity}) Requires that similar images in the retina should correspond to similar unified latent images and similar overlap images,
\item[Consistency radius] (Example \ref{eg:binocular_consistency}) Measures how different the two images are from each other by way of comparing them with the unified and overlap images, and finally
\item[$p$-value] (Example \ref{eg:binocular_pvalue}) Detects when the differences between the images are unlikely to have arisen from a single scene.  When the $p$-value is low, it is unlikely that the observations from each eye could have arisen from a single scene.
\end{description}
In this way, the tower of increasingly stringent consistency checks provides for more complete models, from which stronger inferences can be drawn.

 \section{A worked example: least squares regression}
 \label{sec:linear_example}

 Before building the tower of consistency checks with mathematical formality,
 let us show how diagrammatic modeling works informally first.
 Suppose we wish to find the best fit line $L$ for four points, $A$, $B$, $C$, $D$,
  where
  \begin{equation*}
    A = (0,y_A),
  \end{equation*}
  \begin{equation*}
    B = (1.5,y_B),
  \end{equation*}
  \begin{equation*}
    C = (1.5,y_C),
  \end{equation*}
  and
  \begin{equation*}
    D = (3,y_D).
  \end{equation*}
  By a best fit line $L$, we seek two numerical values, for the slope $m$ and intercept $b$, such that the equation of the line
  \begin{equation}
    \label{eq:line}
    y = mx + b
  \end{equation}
  best approximates the $y$ value for each point, when given its $x$ value.
  For instance, if we consider the point $B$, which has $x=1.5$, we would aim to have
  \begin{equation*}
    y_B \approx 1.5 m + b.
  \end{equation*}
  Put a little more precisely, we would like the quantity
  \begin{equation*}
    | y_B - (1.5 m + b) |
  \end{equation*}
  to be as small as possible.

  Before we attack the problem directly, the perspective advocated in this chapter is that we should consider the variables in play and the dependencies between them.
  For instance, it is natural to suppose that there is an $X$ variable and a $Y$ variable, and that $Y$ depends on $X$ according to Equation \eqref{eq:line}.
  We might draw a picture of that relationship with an arrow
  \begin{equation*}
    \xymatrix{
      X \ar[r] & Y
    }
  \end{equation*}
  This idea is a good one and leads rather directly to \emph{causal modeling} \cite{Pearl_why}.
  On the other hand, since the relationship between $X$ and $Y$ is somewhat ``buried in the arrow'',
  this chapter takes a complementary perspective that unpacks the relationship somewhat more explicitly.
  
  To that end, we suppose that there is a hidden ``latent model'' expressed by Equation \eqref{eq:line} that governs each of the data points we have observed.
  So instead of separating the $X$ and $Y$ variables, think of each point $A$, $B$, $C$, and $D$ as a single observation that depends on the line $L$.
  We can express this idea with a \emph{directed graph} (Definition \ref{def:graph}),
  \begin{equation*}
    \xymatrix{
      &A&\\
      B&L\ar[u]\ar[l]\ar[d]\ar[r]&C\\
      &D&\\
      }
  \end{equation*}
  The idea is that if we were to know the $m$ and $b$ parameters in Equation \eqref{eq:line}, we would know the line $L$,
  and this information determines each of our observations.
  This directed graph is an example of a \emph{span} (Definition \ref{def:span}),
  since it expresses that observable variables are predicted from a latent variable.
  Each arrow starting at $L$ and pointing to a data point represents the process of collecting that data point,
  under the hypothesis that it ``came from'' the line $L$.
  We call each of the $A$, $B$, $C$, $D$, and $L$ the \emph{vertices} of the directed graph,
  and each of the arrows $L \to A$, $L \to B$, $L \to C$, $L \to D$ the \emph{edges} of the directed graph.

  We can add some more information to the vertices and edges,
  so that the actual values of the variables are directly computable.
  Let us start by labeling each vertex with the set of possible values its corresponding variable might take.
  For instance, given that the line $L$ is determined by its slope $m$ and intercept $b$,
  we can represent this as a point in $(m,b) \in \mathbb{R}^2$.
  On the other hand, each of the data points consisted of an $x$ value and a $y$ value.
  It would seem reasonable to label each data point with a point in $\mathbb{R}^2$ as well,
  but it turns out that this is not quite the best choice.
  The $x$ values are better thought of as part of the process that generated each data point as explanatory variables,
  so that the $y$ values are better thought of as the response variables.
  As such, only $y$ values are used as labels on the data points, so they are points in $\mathbb{R}$.
  Thus far, our relabeled graph is
  \begin{equation*}
    \xymatrix{
      && \mathbb{R} &&\\
      \mathbb{R} && \mathbb{R}^2 \ar[u] \ar[ll] \ar[rr] \ar[d] && \mathbb{R} \\
      && \mathbb{R} &&\\
      }
  \end{equation*}
  
  Since we have labeled the vertices of the graph with sets, it is natural to label the edges with functions.
  The domain and range for the function on an edge should be what is specified by the vertices on its ends.
  Each of the four edges have the same domain, namely the set of $(m,b) \in \mathbb{R}^2$.
  Their ranges are also the same set, namely the set of $y \in \mathbb{R}$,
  so to encode the idea that the line is given by Equation \eqref{eq:line},
  it seems like we need to write this equation on each edge.
  That is correct, with one caveat: the $x$ values are ``frozen in'' to each edge.
  That is, the edge $L \to A$ will use $x=0$, whereas the edge $L \to B$ will use $x=1.5$.
  The fully labeled graph is 
  \begin{equation*}
    \xymatrix{
      && \mathbb{R} &&\\
      \mathbb{R} && \mathbb{R}^2 \ar[u]^{b} \ar[ll]_{1.5m+b} \ar[rr]^{1.5m+b} \ar[d]_{3m+b} && \mathbb{R} \\
      && \mathbb{R} &&\\
      }
  \end{equation*}
  To express the transformation from labeling the directed graph with names of variables to sets and functions,
  we call this kind of graph a \emph{sheaf diagram} (Definition \ref{def:sheaf}).

  Given the sheaf diagram, we can show the original regression problem graphically.
  Specifically, we know the values of each of the data points, but not the line.
  \begin{equation*}
    \xymatrix{
      && y_A &&\\
      y_B && (m,b) \ar[u]^{b} \ar[ll]_{1.5m+b} \ar[rr]^{1.5m+b} \ar[d]_{3m+b} && y_C \\
      && y_D &&\\
      }
  \end{equation*}
  This selection of values from each of the corresponding sets of possible values is called an \emph{assignment} (Definition \ref{def:assignment}).
  If we call the assignment $a$,
  we can think of it as a function from the the vertices of the graph to the corresponding sets.
  So, for instance the assignment in the regression problem has $a(A) = y_A$.  
  The regression problem itself then consists of finding the best choice for $a(L) = (m,b)$,
  given that we already know that
  \begin{equation*}
    a(A) = y_A, \; a(B) = y_B, \; a(C) = y_C, \; a(D) = y_D.
  \end{equation*}

  The ``best choice'' for $a(L) = (m,b)$ should be the one that yields the best approximations using Equation \eqref{eq:line}.
  According to classical statistics, the quality of the approximation is measured by taking the sum of squares of the errors at each data point,
  \begin{equation*}
    |b-y_A|^2 + |1.5m+b-y_B|^2 + |1.5m+b-y_C|^2 + |3m+b-y_D|^2.
  \end{equation*}
  This quantity is generalized substantially by the \emph{consistency radius} of the assignment (Definition \ref{def:consistency_radius}).
  A little work shows that the minimum value of the above expression occurs when $m=1$, $b=1$.
  This is one of the most important points of this chapter, that \emph{minimizing consistency radius is a general inference procedure}.

\section{State and sense data}
\label{sec:graphs}

A \emph{directed graph} is a convenient representation for how a model posits that various data are related to one another.
Over the course of this chapter, we will add greater definition to a model,
using a directed graph as a starting point.

The formal definition of a directed graph (below, Definition \ref{def:graph}) posits the existence of two kinds of data: \emph{vertices} and \emph{edges}.
Each vertex represents a datum, which may be an observation, a latent state, or a model parameter (we will defer discussion of types and values of data until Section \ref{sec:sheaves}).
Each edge expresses a dependency between two data, wherein the destination datum depends upon the source datum.
As such, edges ``summarize'' from richer data to poorer data.
Additionally, edges can contain ``frozen in'' parameters governing the dependency.

\begin{definition}
  \label{def:graph}
  A \emph{directed graph $G=(V,E)$} consists of a finite set $V$ of \emph{vertices} and a finite set $E \subseteq V\times V$ of \emph{edges}.
  If $e = (a,b)$ is an element of $E$, we often write $a\to b$ to express the fact that both $a$ and $b$ are vertices,
  and say that $e$ is an \emph{edge from $a$ to $b$}.
  Moreover, $a$ is the \emph{source of $e$}, and $b$ is the \emph{destination of $b$}.
\end{definition}

The definition of a directed graph does \emph{not specify any details} about the vertices or edges,
since that will come later as we add more structure.
However, even the limited information contained in a directed graph is sufficient to identify the presence of spans or cospans in the graph (to be defined in Section \ref{sec:subgraphs}),
which indicate a consistency tests can be performed.

\begin{example} (Overlapping vision)
  \label{eg:binocular_graph}
  Consider binocular vision, by which we simply mean that two eyes are viewing overlapping portions of a scene.
  In some animals, there is a narrow overlap, and three-dimensional views may not be unambiguously derived from vision,
  whereas in others, a more substantial overlap can give rise to a three-dimensional view.
  Therefore, we do not assume immediately that the goal of binocular vision is to result in the perception of depth.
  
  The left and right eye independently collect scene data from the light falling on each retina,
  and these data are then sent to the brain for processing.
  If the two views of the scene from the two eyes are consistent enough,
  the brain can form a coherent image of the scene and can additionally infer the presence of depth in the scene.
  On the other hand, if the two views of the scene are not consistent,
  the phenomenon of \emph{binocular rivalry} occurs.
  Binocular rivalry can produce the sensation of one or the other eye's views dominating, and sometimes the dominating view changes abruptly \cite{Kovacs_1996}.

  A small graph that represents the binocular rivalry problem consists of three vertices: $L$, $R$, and $B$, and two edges $L\to B$ and $R \to B$.
  $L$ represents the view from the left eye and $R$ represents the view from the right eye.
  $B$ represents the view in the overlap region that is seen by both eyes.
  The graph can be expressed conveniently by the diagram
  \begin{equation*}
    \xymatrix{
      L\ar[dr]&&R\ar[dl] \\
      &B&\\
    }
  \end{equation*}

  The model implied by this graph asserts that the two eyes $L$ and $R$ produce sense data that gets compared (by the brain) in the overlap region $B$.
  In Section \ref{sec:sheaves}, the edges will eventually become functional relationships.
  As a result, the edges $L \to B$ and $R \to B$ should not be thought of as \emph{combining} the images from the eyes,
  but rather summarizing the information that can be compared.
  
  A mathematically accurate way to think of the situation is that if we only have the view from $L$ (because the right eye is closed, perhaps),
  then the model asserts that this view determines $B$.
  Moreover, the summary of the scene in $B$ \emph{could} be seen by the right eye $R$ as well, if had been open.
  If observations from both eyes are available, there may be conflict between the eyes if they disagree about $B$,
  resulting in no well-defined view in the overlap region.
  
  No less important, but still hidden in this model, is the idea of a \emph{unified} view of the scene formed by non-rivalrous binocular vision.
  This will become revealed later in Example \ref{eg:binocular_topology}.
\end{example}

When considering binocular vision, the sense data produced by the eyes are fundamentally the same type---visual images.
Often there are situations where a phenomenon is best described as being \emph{multi-modal},
which is to say that several different types of sense data are in play.

\begin{example} (Hidden transmitter)
  \label{eg:fox_graph}
  A concrete case of a multi-modal scene occurs during search-and-rescue operations.
  A radio beacon attached to a stranded individual can be detected by various kinds of sensors that report its relative distance, signal strength, or other location data \cite{Robinson_hunting}.
  While modern GPS-enabled beacons can yield very precise location data on their own (by recording signals from multiple satellites), these data can be inaccurate.
  Moreover, the beacons can become dislodged, forcing rescuers to use simpler technology \cite{Joslyn_bears_2020}.

  We can once again form a graph to represent the situation where a hidden transmitter's signal is detected by several sensors, from which the location of the hidden transmitter can be deduced.
  One such graph is shown below,
  \begin{equation*}
  \xymatrix{
    S_1 && S_2 \\
    & X \ar[ur]\ar[ul]\ar[dl]\ar[dr] &  \\
    S_3 && S_4 \\
    }
\end{equation*}
The location of the hidden transmitter, which is unknown to the sensors but ultimately determines their measurements, is represented by the vertex $X$.
There are four different sensors $S_1$, $S_2$, $S_3$, and $S_4$, each at different locations, and each with different capabilities:
\begin{description}
\item[$S_1$] can measure the bearing (compass direction) to the hidden transmitter,
\item[$S_2$] can measure the signal strength of the hidden transmitter,
\item[$S_3$] can measure both the signal strength and bearing of the hidden transmitter, and
\item[$S_4$] can measure the signal strength of the hidden transmitter.
\end{description}
The presence of four edges $X \to S_k$, for $k=1, \dotsc, 4$, implies that if the hidden transmitter's location were to be known,
then we could use physical principles to predict the measurements from each of the four sensors.
\end{example}

Note well that unlike the graph in the overlapping vision example (Example \ref{eg:binocular_graph}), in which the edges point ``into'' the vertex $B$,
in the graph in the hidden transmitter example (Example \ref{eg:fox_graph}), the edges point ``away'' from the vertex $X$.
The difference between these two graph structures will play an important role in what follows.
When edges point ``into'' a vertex, this yields an opportunity for a data consistency test,
whereas when the edges point ``away'' from a vertex, this is an opportunity to specify a hypothesis about how data sources combine into a unified picture.
These are not the only kind of structures within graphs; two more are \emph{paths} and \emph{loops}.

\begin{definition}
  In a directed graph $G=(V,E)$,
  a sequence $(v_1 \to v_2), (v_2 \to v_3), \dotsc, (v_{n-1} \to v_n)$ of edges is called a \emph{path from $v_1$ to $v_n$},
  which is to say that the destination of each edge is the source of the next edge in the sequence.
  A path from $v$ to $v$ (starting and ending at the same vertex) is called a \emph{loop}.
  A loop consisting of a single edge $v\to v$ is called a \emph{self-loop}.
  If the only loops in a directed graph are self-loops\footnote{Self loops are not required for acyclicity, but will be implied for transitive closure, to be defined later in Definition \ref{def:poset}.}, the graph is said to be \emph{acyclic}.
\end{definition}

Since edges express dependency, the presence of a loop expresses a feedback relationship between several different data.
If $a$ and $b$ are both part of the same loop, then the model expresses that both $a$ depends on $b$, and $b$ depends on $a$.
This \emph{feedback} relationship is entirely possible, especially if data consist of timeseries and the dependency contains a time lag,
for instance.

On the other hand if the dependency conditions are to be interpreted as being instantaneous,
acyclicity is a consistency check on the model represented by a directed graph.
While it is not impossible or undesirable for a model to contain loops,
models described by acyclic directed graphs are easier to analyze.

The previous examples showed that small graphs can be useful for identifying structure in problems.
The graph for a given problem can become quite large, as Example \ref{eg:path_planning} shows.
This is especially useful when it reveals structure that would otherwise be hidden.

\begin{example} (Road trip)
  \label{eg:path_planning}
  Imagine planning a road trip along the northeastern coast of North America, starting at Halifax, Nova Scotia, Canada and ending at Philadelphia, Pennsylvania, USA.
  Although one would naturally consult a map, one might consider building a graph that summarizes the most relevant waypoints along the journey.
  In such a graph, the vertices are cities, and the edges are the driving routes from one city to the next.
  Taking inspiration from the idea that vertices should represent variables and edges are dependencies between them,
  we can imagine that the main consideration is the total accumulated driving time along the journey.
  Each edge then supplies a prediction of what the driving time between the two waypoints it connects will be.

  \begin{table}
    \begin{center}
      \caption{Cities in the northeastern part of North America, used in Example \ref{eg:path_planning}}
      \label{tab:path_planning}
      \begin{tabular}{|l|l|l|l|}
        \hline
        Abbrev. & City name & State/Province & Country\\
        \hline
        $HFX$  & Halifax & Nova Scotia & Canada\\
        $BAN$ & Bangor & Maine & USA \\
        $AUG$ & Augusta & Maine & USA \\
        $PME$ & Portland & Maine & USA\\
        $PNH$ & Portsmouth & New Hampshire & USA\\
        $LOW$ & Lowell & Massachusetts & USA\\
        $BOS$ & Boston & Massachusetts  & USA \\
        $WOR$ & Worcester & Massachusetts  & USA\\
        $NHCT$& New Haven & Connecticut & USA\\
        $NLCT$& New London & Connecticut & USA\\        
        $PRD$ & Providence & Rhode Island & USA\\
        $PHL$ & Philadelphia & Pennsylvania & USA\\
        $SCR$ & Scranton & Pennsylvania & USA\\
        $NYC$ & New York & New York & USA\\
        \hline
      \end{tabular}
    \end{center}
  \end{table}

  After consulting a map with estimated driving times, one might obtain the graph shown below (see Table \ref{tab:path_planning} listing the names of the cities involved),
  where the edges are labeled driving time (in hours) between cities.
  \begin{equation*}
    \xymatrix{
      HFX \ar[r]^{3} & BAN \ar[r]^{1} & AUG\ar[r]^{1} & PME \ar[r]^{1} & PNH \ar[r]^{1} \ar[d]_{1} & LOW \ar[d]^{0.7} \ar[dl]_{0.7} \\
                &            &           &            & BOS \ar[dr]^{1.1}\ar[r]^{1} & WOR \ar[d]^{0.7} \\
                &            & SCR \ar[dl]_{2}\ar[d]^{2} & NHCT \ar[l]_{3} \ar[dl]^{2}& HAR \ar[d]^{0.8} \ar[l]^{0.7}& PRD \ar[l]_{1.5}\ar[dl]^{1} \\
                & PHL       & NYC\ar[l]^{1.7}          &                             & NLCT \ar[lu]^{0.8} & \\
      }
  \end{equation*}

  Certain features that would not be obvious from a table of distances become visible in the graph.
  For instance, certain cities are decision points, namely $PNH$, $PRD$, $NHCT$.
  One must pass through each of these cities, and moreover once one has reached one of these locations, there is a choice about how to proceed.

  It is also important to recognize the difference between a path and an edge.
  While there is a path $HFX \to \dotsb \to PHL$, there is no edge $HFX \to PHL$.
  The distinction lies in the following intuition: the presence of a path means that it is possible to start at $HFX$ and ultimately arrive at $PHL$,
  whereas an edge $HFX \to PHL$ would imply that one can travel \emph{directly} from $HFX$ to $PHL$ without making intermediate stops.
  Sometimes this distinction is useful, and sometimes it is not!  
\end{example}

\section{Useful structures within graphs}
\label{sec:subgraphs}

The road trip example (Example \ref{eg:path_planning}) suggests that the distinction between a dependency expressed by an edge versus a path can be an important one.
Conversely, it is useful to consider graphs in which this distinction vanishes, called \emph{transitively closed graphs}.

\begin{definition}
  \label{def:poset}
  A directed graph is said to by \emph{transitively closed} if whenever there is a path from $a$ to $b$,
  there is also an edge $a\to b$.

  A \emph{partially ordered set} is a set $P$ and a relation $\le$,
  which means that
  \begin{description}
  \item[Reflexivity] If $a \in P$, then $a\le a$ holds.
  \item[Antisymmetry] If $a \le b$ and $b \le a$ both hold, then $a=b$.
  \item[Transitivity] If $a\le b$ and $b\le c$ both hold, then $a\le c$.
  \end{description}
\end{definition}

Evidently, the graphs in the overlapping vision example (Example \ref{eg:binocular_graph}) and the hidden transmitter example (Example \ref{eg:fox_graph}) are transitively closed,
since there is no opportunity for trouble.
In contrast, the graph in the road trip example (Example \ref{eg:path_planning}) is not transitively closed.

\begin{corollary}
  \label{cor:poset}
  Suppose that $G=(V,E)$ is an directed acyclic graph that is also transitively closed, and every vertex is part of a self-loop.
  Taking the presence of an edge $a\to b$ as a relation $a \le b$ between vertices makes the set of vertices $V$ into a partially ordered set.
\end{corollary}

The reader is warned that drawing transitively closed graphs quickly gets visually cluttered.
If we were to add all the edges to the graph in the road trip example (Example \ref{eg:path_planning}) needed to make it transitively closed,
the structure of the decision points $PNH$, $PRD$, $NHCT$ would become obscured.
Therefore, we will generally avoid drawing all of the edges needed to make a graph transitively closed,
preferring instead to merely state that the graph in question has all necessary edges for this to be true.
In this way, we make use of the fact that graphs can contain others as \emph{subgraphs},
which is formalized by the next definition.

\begin{definition}
  Suppose that $G_1=(V_1,E_1)$ and $G_2=(V_2,E_2)$ are two directed graphs,
  in which $V_1 \subseteq V_2$ (every vertex of $G_1$ is also vertex of $G_2$) and $E_1 \subseteq E_2$ (every edge of $G_1$ is an edge of $G_2$).
  We say that $G_1$ is a \emph{subgraph} of $G_2$.
\end{definition}

Studying subgraphs of specific forms is a useful way to characterize the relationships expressed in a larger graph.
Two kinds of subgraphs are especially important in understanding the dependency relationships between data in a model.

First, suppose a collection of variables $A_1, \dotsc, A_n$ all share a common source $B$.
This is expressed in a directed graph by a collection of edges $B \to A_i$, so the whole graph might look like
\begin{equation*}
  \xymatrix{
    \text{(Span)}&& B \ar[dr] \ar[dl] & \\
    &A_1 & \dotsb & A_n \\
    }
\end{equation*}
This kind of graph is called a \emph{span}.
As a very small example, $B$ could be the financial bank ledger of a store, and $A_1$, $\dotsc$, $A_n$ could represent the records of each cash register.
If we have access to the bank ledger, then this completely determines what we would expect to see in each cash register's record as well,
because the ledger should contain all of the store's transactions in their entirety.

We interpret a span as a building block for many hypotheses we might make about a model.
If $B$ is known then we can predict all of the values of the $A_i$ from the value of $B$.
This is typically called a \emph{forward problem}.
Conversely, if $B$ is not observable but the $A_i$ are, it is common to say that $B$ is a \emph{latent} variable,
and the problem of finding values of $B$ given values of the $A_i$ is called an \emph{inverse problem}.

If we reverse the arrows in a span, we obtain a graph that looks like
\begin{equation*}
  \xymatrix{
    \text{(Cospan)}&& B & \\
    &A_1 \ar[ur] & \dotsb & A_n \ar[ul]\\
    }
\end{equation*}
This kind of graph is called a \emph{cospan}.
If $A_1$, $\dotsc$, $A_n$ represent the addresses on each piece of mail delivered to a residence, then $B$ could represent the postal code associated to that residence.
Each address should contain the postal code, and moreover, all of the mail delivered to the residence should have the same postal code.

We can interpret a cospan as saying that all of the $A_i$ can be used to predict $B$ as a common summary.
If we observe values for all of the $A_i$, and they all predict the same value for $B$, we say that the values are \emph{consistent}.
What is more interesting is that if we know some---but not all---consistent values for the $A_i$ variables,
then a cospan places a constraint on the possible consistent values for the remaining $A_i$ variables.
This modeling behavior can be very counter-intuitive! 

In a very concrete sense, \emph{spans build complex hypotheses; cospans test them}.
With these ideas in mind, the formal definitions of span and cospan should not be surprising.

\begin{definition}
  \label{def:span}
  Suppose that $V = \{a,b,c\}$ are three vertices of a graph.
  A subgraph whose edges are $\{a \to b, a\to c\}$ is called a \emph{span}.
  Dually, a subgraph whose edges are $\{b \to a, c\to a\}$ is called a \emph{cospan}.
  We call the vertex $a$ the \emph{center} of the span or cospan.
\end{definition}

We have already seen a cospan in the overlapping vision example (Example \ref{eg:binocular_graph}) and a span in the hidden transmitter example (Example \ref{eg:fox_graph}).
Moreover, the road trip example (Example \ref{eg:path_planning}) contains several examples of each.
As the rest of the chapter will show, nearly every statistical inference problem can be viewed as alternating spans and cospans.
Sometimes they naturally come in pairs, as Example \ref{eg:lightning_graph} below shows.

\begin{example}(Lightning)
  \label{eg:lightning_graph}
  When a lightning strike occurs nearby, the flash from the bolt arrives nearly instantaneously, but the rumble of thunder arrives some time later.
  The apparent separation of these two events in time is due to the fact that the speed of sound is much slower than the speed of light.
  If two observers at different locations with synchronous clocks experience the same lightning event,
  they will report the same flash time (within less than a millisecond) but potentially different thunder times,
  because the speed of light is much faster than the speed of sound.

  Suppose that the lightning strike location and time are represented by $X$,
  and the two observers are represented by $A$ and $B$.
  Both of their observations are determined by $X$ through physical laws.
  While they will hear the sound differently,
  both of the observers will agree on the flash time $T$.
  This can be represented by the following directed graph,
  \begin{equation*}
    \xymatrix{
      &X\ar[dr]\ar[dl]&\\
      A\ar[dr]&&B\ar[dl]\\
      &T&\\
      }
  \end{equation*}
  This graph has a span centered on $X$, because $X$ determines both $A$ and $B$.
  It has a cospan centered on $T$, because $A$ and $B$ should share the same flash time $T$.
  It is not expected that their records of the thunder time will be similarly comparable.
  Of course, if their clocks are inaccurate, they will disagree about $T$ as well.
\end{example}

\begin{example} (Regression models)
  \label{ex:competing_models}
  A common problem in statistics is to find a simple equation that best fits a set of data points.
  Intuitively, fitting a line affords less freedom than fitting a curve, so what counts as a best-fit line or a best-fit curve could be different.
  Nevertheless, both can be tested against the data.
  
  Suppose we have five observations $O_1=(x_1,y_1)$, $O_2=(x_2,y_2)$, $O_3=(x_3,y_3)$, $O_4=(x_4,y_4)$, and $O_5=(x_5,y_5)$,
  which correspond to points in the plane.
  We could imagine a curve model $M_k$ for these data points which defines a polynomial of degree $k$,
  \begin{equation*}
    y = a_0 + a_1 x + \dotsc a_k x^k.
  \end{equation*}
  If we substituted $x_i$ into such an equation, we will obtain a \emph{prediction} $\widehat{y_i}$.
  Ideally, the predictions fit the data, which means that we hope that $\widehat{y_i}=y_i$, for $i=1, \dotsc, 5$.

  Each model, say $M_1$, corresponds to a single span:
  \begin{equation*}
    \xymatrix{
      M_1 \ar[d]\ar[dr]\ar[drr]\ar[drrr]\ar[drrrr]& & & &\\
      O_1 & O_2 & O_3 & O_4 & O_5\\ 
    }
  \end{equation*}
  Structurally, we can express the relationships between all of the models together and the observations as a single graph,
  \begin{equation*}
    \xymatrix{
      M_1 \ar[d]\ar[dr]\ar[drr]\ar[drrr]\ar[drrrr]& M_2 \ar[dl]\ar[d]\ar[dr]\ar[drr]\ar[drrr]& M_3 \ar[dll]\ar[dl]\ar[d]\ar[dr]\ar[drr]& M_4 \ar[dlll]\ar[dll]\ar[dl]\ar[d]\ar[dr]& \dotsb\\
      O_1 & O_2 & O_3 & O_4 & O_5\\ 
      }
  \end{equation*}
  Since there is a span centered at each $M_k$ in the graph, the graph expresses the fact that each model predicts all of the observations.
  Each model $M_k$ is a hypothesis about how the data points are placed on the plane.
  
  Moreover, the observations are what they are, and of course need not agree with the predictions of any of the models.
  This fact is expressed in the graph by the presence of a cospan centered on each observation.
  Briefly, each data point corresponds to a specific test we can run on the models.
\end{example}

\begin{remark}
  Cospans are related to---but not exactly identical to---Judea Pearl's notion of a \emph{collider} in a causal diagram \cite{Pearl_why}.
  The most significant difference is that in this chapter (in Section \ref{sec:sheaves}) we promote edges in the graph to become \emph{functional} relationships,
  whereas the edges in Pearl's causal diagrams are \emph{conditional probability} relationships.
  
  A cospan like
  \begin{equation*}
    \xymatrix{
      a \ar[r] & b &  c \ar[l]
    }
  \end{equation*}
  means that both $a$ and $c$ \emph{individually} determine the value of $b$.
  However, in a causal diagram, the same picture implies that $a$ and $c$ \emph{jointly} determine the value of $b$.

  There is a clear way to translate between these two perspectives.
  Specifically, the causal diagram can be systematically translated into a larger directed graph that exposes functional relationships \cite{Robinson_DSEM_2025},
  by replacing each cospan in the causal diagram with a diagram of the form
  \begin{equation*}
    \xymatrix{
      a & a\times c \ar[l] \ar[d] \ar[r] & c \\
       & b &\\
      }
  \end{equation*}
  where $a\times c$ represents the set of ordered pairs and the arrows in the top row select one or the other variable.
\end{remark}

\section{Topology: Locality of data}
\label{sec:topology}

What is the domain of validity for different parts of a model?
Given a graph of dependency relationships,
there are various opportunities for building hypotheses and tests.
Some of these are automatic (the graph forces them upon us),
whereas others are optional (depending on how broadly applicable we wish our models to be, or how stringent we wish our tests to be).
This flexibility appears concretely in whether we demand (on principle) that the graph be transitively closed.
A topology removes the ambiguity about these choices.
It is a consistency check on locality assumptions, from which spans and cospans---and hence hypotheses and tests---can be derived.

\begin{definition}
  \label{def:topology}
   Suppose $X$ is a set.
   A \emph{topology} on $X$ is a collection $\col{T}$ of subsets of $X$, such that
   \begin{enumerate}
   \item The empty set $\emptyset$ is in $\col{T}$.
   \item The set $X$ is in $\col{T}$.
   \item $\col{T}$ is closed under arbitrary unions, which means that if $\col{U} \subseteq \col{T}$ is any subcollection, then its union $\cup \col{U}$ is an element of $\col{T}$.
   \item $\col{T}$ is closed under finite intersections, which means that if $U$ and $V$ are elements of $\col{T}$, then their intersection $U \cap V$ is in $\col{T}$.
   \end{enumerate}

   The pair $(X,\col{T})$ is called a \emph{topological space}.  Each element of $\col{T}$ is called an \emph{open set} of the topological space.
\end{definition}

\begin{example}
  \label{eg:simple_topology}
  Let $X$ be the set $\{1,3,7\}$.  Among several possibilities, the collection $\col{T}$ forms a topology when defined by
  \begin{equation*}
    \col{T} = \{\emptyset, \{1,3\}, \{7\}, \{1,3,7\} \}.
  \end{equation*}
  Note that in this topology, $1$ and $3$ cannot be separated.
\end{example}

 Although stated in set-theoretic language, Definition \ref{def:topology} is closely related to the graph-based approach from the previous sections.
 If we consider the elements of a topology $\col{T}$ on a set $X$ as vertices,
 and the superset\footnote{Although a subset relationship might feel more natural, and would result in a mathematically equivalent formulation, it is opposite of the standard formulation when one ultimately builds sheaves on topological spaces.} relationship $U \supseteq V$ to correspond to an edge $U \to V$,
 the topology corresponds to a graph.
 We will carry this idea forward throughout the chapter: open sets in the topology will correspond to vertices and to observations, and superset relationships correspond to edges and to dependencies.
 
 Because $A \cup B \supseteq A$ and $A \cup B \supseteq B$,
 each union corresponds to a span,
 \begin{equation*}
   \xymatrix{
     A & A \cup B \ar[r] \ar[l] & B\\
     }
 \end{equation*}
 Moreover, notice that because $A \supseteq A \cap B$ and $B \supseteq A \cap B$,
 this corresponds to a cospan,
 \begin{equation*}
   \xymatrix{
     A \ar[r] & A\cap B & B \ar[l] \\
     }
 \end{equation*}
 Intuitively, Definition \ref{def:topology} states that in the graph of open sets for a topology, all reasonable opportunities for spans and cospans exist.

 Conversely, we can obtain a topology from a partially ordered set in a natural way.
 Recalling that directed acyclic graphs naturally yield partially ordered sets (see Corollary \ref{cor:poset}),
 this topology reflects the structure of the dependencies between variables represented in the graph.
 In many applications, this topology is all that is needed to specify all of the sub-models and tests within a larger model.

 Given a set $P$ with a partial order $\le$ and an element $x \in P$,
 we can consider all paths in the directed acyclic graph for $P$ that start at $x$ and end somewhere else.
 The set of possible destinations with source $x$ is
 \begin{equation*}
   \{y \in P : x \le y \}.
 \end{equation*}
 The \emph{Alexandrov topology} $Alex(P,\le)$ is defined by taking every possible union of these sets.
 That is, we define 
 \begin{equation*}
   \bigcup_{x \in S} \{y \in P : x \le y \}
 \end{equation*}
 to be an open set, where $S \subseteq P$ arbitrary.

 \begin{lemma}(see \cite{Barmak_2011})
   \label{lem:alexandrov}
   The Alexandrov topology is a topology.
 \end{lemma}
 
 The point of Lemma \ref{lem:alexandrov} is that one can manufacture \emph{all relevant} spans and cospans for use in building and testing hypotheses,
 and as a result, Lemma \ref{lem:alexandrov} is sometimes called the \emph{Alexandrov construction}.
 Alexandrov construction is valuable because it reveals places for comparison that were implicit in the original dependency graph.
 
 In some cases, the construction does not reveal any new spans or cospans.
 This is what happens when we apply the Alexandrov construction to a pre-existing span.

 \begin{example} (Search and rescue)
  \label{eg:titan_graph}
  Consider the unfortunate incident of the submersible \emph{Titan}.
  While on an underwater expedition to the wreck of the \emph{Titanic},
  the hull of the \emph{Titan} collapsed, resulting in an implosion that was detected by various underwater acoustic (sound) sensors.
  Based upon the arrival times of the sound wave pulse, rescuers attempted \cite{titan_press_release} to find the destroyed \emph{Titan}.
  They ultimately succeeded in finding the wreckage several days later after an extensive search of the area \cite{titan_report}.
  
  This situation can be modeled as a directed graph in which a vertex $X$ represents the time and location of implosion,
  and (say) six sensors $S_1$, $S_2$, $S_3$, $S_4$, $S_5$, and $S_6$ recorded the sound pulse's arrival time.
  Since the true (but unknown) time and location of the implosion determines the pulse arrival times at each of the sensors,
  the structure of the directed graph could be as follows,
  \begin{equation*}
    \xymatrix{
      S_1 & S_2 & S_3 \\
      &X\ar[ul]\ar[u]\ar[ru]\ar[dl]\ar[d]\ar[dr] & \\
      S_4 & S_5 & S_6 \\
      }
  \end{equation*}
  In this model, each sensor $S_1$, $\dotsc$, $S_6$ has a known location, but since these are not in dispute they do not show up as vertices in the graph.

  The Alexandrov topology associated to this graph consists of all unions of the following sets,
  \begin{equation*}
    \emptyset, \{S_1\}, \{S_2\}, \{S_3\}, \{S_4\}, \{S_5\}, \{S_6\}, \{X,S_1,S_2,S_3,S_4,S_5,S_6\}. 
  \end{equation*}
  In total, there are $2^6+1 = 65$ open sets in the topology.
  Note that the edges of the graph tell you that if you include $X$, all of its descendants ``come along'' as the definition of topology is now in play.
  This tells us that the moment $X$ is known, then predictions are available for \emph{all} sensors, regardless of whether they were observed or not.
  
  From a modeling perspective, most of these open sets are uninformative, because the spans they imply are already present in the original graph.
  For instance,
  \begin{equation*}
    \xymatrix{
      \{S_1\} & \{S_1, S_2\} \ar[r]\ar[l] & \{S_2\}\\
      }
  \end{equation*}
  does not help us find the unknown location of the \emph{Titan} because $X$ is not an element of any of the three sets.
  The only informative span is the one centered on $\{X,S_1,S_2,S_3,S_4,S_5,S_6\}$,
  yet this span corresponds to the original graph anyway.

  Although the location of the \emph{Titan} could be inferred from four time observations, say $S_1$, $S_2$, $S_3$, $S_4$,
  this would imply a different topology than the one we have defined.
  In particular, such a topology would have 
  \begin{equation*}
    \{X, S_1, S_2, S_3, S_4\}
  \end{equation*}
  as an open set.
  This set is not open in the topology we obtained from the graph.
 \end{example}

 We can also iterate the construction, starting with a graph to create a topology, and from that topology obtain a new graph structure that includes these hidden spans.
 Applying the iterated Alexandrov construction to a cospan can reveal a ``hidden'' span.
 Revealing a hidden span is useful for the binocular rivalry case, as highlighted in the next example. 

 \begin{example} (Overlapping vision topology)
  \label{eg:binocular_topology}
  Recall that in the overlapping vision example (Example \ref{eg:binocular_graph}), we started with the graph with one cospan,
  \begin{equation*}
    \xymatrix{
      L\ar[dr]&&R\ar[dl] \\
      &B&\\
    }
  \end{equation*}
  If we form the Alexandrov topology for this graph, all open sets are unions of the following five sets,
  \begin{equation*}
    \emptyset, \{B\}, \{L,B\}, \{R,B\}, \{L,R,B\}.
  \end{equation*}

  We can organize these into a new directed graph, where an edge corresponds to the superset relation.
  To start, we replace each vertex in the original graph with a set containing itself and all of its descendants.
  For instance, since $\{L,B\} \supset \{B\}$, we have an edge $\{L,B\} \to \{B\}$.
  The resulting graph for the Alexandrov topology is
  \begin{equation*}
    \xymatrix{
      &\{L,R,B\} \ar[dr] \ar[dl] \ar[dd] \ar@/r{10pc}/[ddd]& \\
      \{L,B\}\ar[dr]\ar[ddr] & & \{R,B\}\ar[dl]\ar[ddl]\\
      &\{B\}\ar[d]&\\
      &\emptyset&\\
      }
  \end{equation*}
  Notice that each of the vertices $L$, $R$, $B$ in the original graph are present in the Alexandrov topology graph.
  Namely $L$ corresponds to $\{L,B\}$, $R$ corresponds to $\{R,B\}$, and $B$ corresponds to $\{B\}$.
  The edges in the original graph are also present in the Alexandrov topology graph.
  To simplify notation a bit for this specific example, we can ignore $\emptyset$ and rename each of the vertices to obtain
  \begin{equation*}
    \xymatrix{
      & V \ar[dr]\ar[dl]\ar[dd]\\
      L \ar[dr]  & & R \ar[dl]\\
      &B&\\
      }
  \end{equation*}
  
  The Alexandrov topology has revealed the presence of a hidden span (centered on $V$) lurking within our original model!
  Since spans build complex hypotheses, we should expect to see some additional modeling opportunities.
  Indeed, the open set $V=\{L,R,B\}$ has two rather different interpretations.
  In one interpretation, which is \emph{causal},  $V$ represents the ``actual'' scene.
  Although it cannot be sensed directly, it determines what each eye sees.
  This determination is encoded in the two edges $V \to L$ and $V\to R$.
  In the other interpretation, which is \emph{explanatory}, $V$ represents the ``reconstructed'' scene within the mind.
  
  The difference between the two interpretations is subtle.
  The causal interpretation states that ``because the scene $V$ is this way, it causes the left eye $L$ to perceive the following sense data...''
  The explanatory interpretation states that ``the scene $V$ being this way explains why the left eye $L$ perceived the sense data it did.''
  At this point in the chapter, we cannot distinguish between these two interpretations.
  While the edges $V \to L$ and $V \to R$ might have different mathematical functional expressions (Example \ref{eg:binocular_sheaf} gives two such),
  this need not be the case.
  It is only when we discuss \emph{assignment supports} (Definition \ref{def:assignment}) that the interpretations necessarily differ,
  as shown in Example \ref{eg:binocular_section}.
\end{example}

If a span and a cospan are already related to each other, then the iterated Alexandrov construction will not reveal anything else new. 

\begin{example}(Lightning topology)
  \label{eg:lightning_topology}
  Recall the lightning example (Example \ref{eg:lightning_graph}),
  which considered the graph
  \begin{equation*}
        \xymatrix{
      &X\ar[dr]\ar[dl]&\\
      A\ar[dr]&&B\ar[dl]\\
      &T&\\
      }
  \end{equation*}
  The Alexandrov topology is
  \begin{equation*}
    \emptyset, \{T\}, \{A,T\}, \{B,T\}, \{X,A,B,T\}.
  \end{equation*}
  Although $A$ and $B$ do not completely determine the location on their own, they do agree on $T$, the flash time.
  It happens that $A$ and $B$ can constrain the strike location in $\mathbb{R}^2$ to one or two locations by intersecting circles.
  If there are more sensors, like the search and rescue example (Example \ref{eg:titan_graph}), then one can constrain the location completely.
\end{example}

\section{Sheaves: Stating a local-to-global hypothesis}
\label{sec:sheaves}

Once again, our perspective is that spans build models and cospans test them.
For a given modeling framework, a topology characterizes all of the spans and cospans that are relevant to a variable dependency graph.
A \emph{sheaf} fully describes a model by adding the variables themselves and their functional relationships to the topology (and hence the variables, observations, and dependency relationships).
As a result, the sheaf specifies all of the relevant functional dependencies between variables in a model.
With these in hand, we can test these functional dependencies for internal consistency of the model.

The complexity of the sheaf definition below simply amounts to redrawing the original graph with richer labels.
When we redraw the graph with all the data required by Definition \ref{def:sheaf},
we emphasize this fact by saying that we have drawn a \emph{sheaf diagram} rather than a graph.
We use special notation to reflect this as well,
\begin{itemize}
\item The name of the sheaf uses a single capital script character, for instance $\shf{S}$.
\item If $v$ is a vertex, the label within the sheaf diagram uses ``$(v)$'' after the name of the sheaf, like $\shf{S}(v)$, suggesting that part of the definition of the sheaf specifies a function whose domain is the set of vertices.
\item If $e$ is an edge $a\to b$, the label within the sheaf diagram uses ``$(e)$'' or ``$(a \to b)$'' after the name of the sheaf, like $\shf{S}(e)$ or $\shf{S}(a\to b)$.  If we use the latter notation, there is no confusing the labels of vertices of edges.  
\end{itemize}

With this in mind, the formal definition of a sheaf follows.

\begin{definition}
  \label{def:sheaf}
  Suppose that $G=(V,E)$ is a directed graph.
  A \emph{sheaf $\shf{S}$ of sets on $G$} (briefly a \emph{sheaf on $G$}) consists of the following information:
  \begin{description}
  \item[Stalks:] To each vertex $v$ in $V$, the sheaf $\shf{S}$ assigns a set $\shf{S}(v)$, called the \emph{stalk at $v$},
  \item[Restrictions:] To each edge $e = (v_1 \to v_2)$ in $E$, $\shf{S}$ assigns a function $\shf{S}(e) : \shf{S}(v_1) \to \shf{S}(v_2)$, called the \emph{restriction on $e$},
  \end{description}
  such that if there are multiple paths $s = v_1 \to \dotsb \to v_n = d$ starting at a vertex $s$ and ending at a vertex $d$, then the composition of functions
  \begin{equation*}
    \left(\shf{S}(v_{n-1} \to v_n)\right) \circ \dotsb \circ \left(\shf{S}(v_1 \to v_2)\right)
  \end{equation*}
  depends only on $s$ and $d$, so that we may write $\shf{S}(s \to d)$ without ambiguity.
\end{definition}

The condition about path independence is usually stated as, ``the diagram of a sheaf is commutative,'' or ``a sheaf diagram commutes.''
Commutativity of the sheaf diagram is a consistency check on the structure of the model.
If there is a dependency between variables, then commutativity ensures that there is a well defined way to functionally predict one from the other.

\begin{example}
  \label{eg:triangle_example}
  Suppose we start with a small directed graph
  \begin{equation*}
    \xymatrix{
      &C\ar[dr]&\\
      A\ar[rr]\ar[ur]&&B\\
      }
  \end{equation*}
  We can build a sheaf $\shf{S}$ on this graph by specifying the stalks and the restrictions.
  Formally, the definition requires three stalks, on $A$, $B$, and $C$, and three restrictions, on $A\to B$, $A\to C$, and $C \to B$.
  We have considerable freedom in specifying these data.
  For instance, we might define the stalk on $A$ by
  \begin{equation*}
    \shf{S}(A) = \mathbb{R}.
  \end{equation*}
  Let us make the same definition for the other stalks, too, so that
  \begin{equation*}
    \shf{S}(B) = \mathbb{R},
  \end{equation*}
  and
  \begin{equation*}
    \shf{S}(C) = \mathbb{R}.
  \end{equation*}
  The sheaf diagram has the same shape as the original graph, so we can simply write the restrictions as labels on the edges of the graph.
  Here is one possible choice of restrictions that satisfies the definition of a sheaf,
  \begin{equation*}
    \xymatrix{
      &\mathbb{R}\ar[dr]^{g(x)=x+17}&\\
      \mathbb{R}\ar[rr]_{f(x)=2x+17}\ar[ur]^{h(x)=2x}&&\mathbb{R}\\
      }
  \end{equation*}
  Unpacking this into the notation specified in Definition \ref{def:sheaf}, the restriction on $A\to B$ is the function $f$, which we can write as
  \begin{equation*}
    \shf{S}(A\to B) = f,
  \end{equation*}
  or if we want to express it more fully, 
  \begin{equation*}
    \left(\shf{S}(A\to B)\right)(x) = f(x) = 2x+17.
  \end{equation*}

  Finally, the commutativity condition is that the restriction along $A\to B$ must be the composition of the restrictions along $A\to C$ and $C\to B$,
  expressed by the equation
  \begin{equation*}
    f(x) = g(h(x)) = (g \circ h)(x).
  \end{equation*}
  Notice that this holds because
  \begin{equation*}
    \begin{aligned}
      f(x) &= 2x + 17\\
      &= h(x) + 17\\
      &= g(h(x)).\\
    \end{aligned}
  \end{equation*}
  
\end{example}

In some cases, the commutativity constraint is trivially satisfied.

\begin{example}(Search and rescue sheaf)
  \label{eg:titan_sheaf}
  We can build a sheaf $\shf{T}$ for the search and rescue example (Example \ref{eg:titan_graph}) by labeling each vertex with the set of possible values each variable can take.
  The edges of the graph are promoted to being functions between these sets.
  Each of the times recorded by the sensors are single real numbers, so this means that $\shf{T}(S_k) = \mathbb{R}$.
  The submersible's location at the time of its implosion is given by four real numbers---three for position $(x,y,z)$ and one for implosion time $t$---so we should define $\shf{T}(X) = \mathbb{R}^4$.
  Schematically, the sheaf will have a diagram like the following,  
  \begin{equation*}
    \xymatrix{
      \shf{T}(S_1)=\mathbb{R} & \shf{T}(S_2)=\mathbb{R} & \shf{T}(S_3)=\mathbb{R} \\
      &\shf{T}(X)=\mathbb{R}^4 \ar[ul]\ar[u]\ar[ru]\ar[dl]\ar[d]\ar[dr] & \\
      \shf{T}(S_4)=\mathbb{R} & \shf{T}(S_5)=\mathbb{R} & \shf{T}(S_6)=\mathbb{R} \\
      }
  \end{equation*}
  As noted above, the graph structure of the sheaf diagram is exactly the same as the graph in the search and rescue example (Example \ref{eg:titan_graph}).
  Each of the vertices in the original graph are now labeled with the corresponding stalks of $\shf{T}$. 

  We still have to define the restrictions.
  From the structure of the diagram, the path independence condition in Definition \ref{def:sheaf} is satisfied automatically,
  since there is at most one path between any pair of vertices.    
  For simplicity, let us assume all of the sensors are surface-based sensors, so the $z$ coordinates of their locations are all zero.
  If sensor $k$ is located at $(x_k,y_k,0)$, then the arrival time of the implosion's sound at that sensor is captured by the restrictions of $\shf{T}$,
  \begin{equation*}
    \left(\shf{T}(X \to S_k)\right)(x,y,z,t) = t + \frac{\sqrt{(x-x_k)^2 + (y-y_k)^2 + z^2}}{v},
  \end{equation*}
  where $v$ is the speed of sound through the water.
  In other words, $\shf{T}(X \to S_k)$ specifies precisely what sound pulse arrival time sensor $S_k$ would have measured, if the implosion were to have occurred at location $(x,y,z)$ at time $t$.
  The restrictions thereby capture the physical modeling assumptions we make about sound transmission through the water.
\end{example}

Of course, the commutativity condition is not always trivial, as the next example shows.

\begin{example}(Lightning sheaf)
  \label{eg:lightning_sheaf}
  Recall the lightning example (Example \ref{eg:lightning_graph}), about two observers experiencing the same lightning flash.
  If we build a sheaf $\shf{L}$ to represent this problem setting,
  it should be clear that the stalk for $X$ should represent the location and time of the lightning strike.
  Therefore, $\shf{L}(X) = \mathbb{R}^2 \times \mathbb{R} \cong \mathbb{R}^3$.
  Each observer records two times, namely the flash time and the thunder time, so $\shf{L}(A) = \shf{L}(B) = \mathbb{R}^2$.
  Finally, the flash time consists of one measurement, so $\shf{L}(T) = \mathbb{R}$.
  The sheaf diagram again matches that of the original graph structure.
  \begin{equation*}
    \xymatrix{
      &\mathbb{R}^3 \ar[dr] \ar[dl]&\\
      \mathbb{R}^2\ar[dr]&&\mathbb{R}^2\ar[dl]\\
      &\mathbb{R}&\\
      }
  \end{equation*}
  The top two restrictions in the diagram are very similar to those in the search and rescue sheaf example (Example \ref{eg:titan_sheaf}),
  \begin{equation*}
    \left(\shf{L}(X \to A)\right)(x,y,t) = (t,t+ 1/v \sqrt{(x-x_A)^2+(y-y_A)^2}),
  \end{equation*}
  \begin{equation*}
    \left(\shf{L}(X \to B)\right)(x,y,t) = (t,t+ 1/v \sqrt{(x-x_B)^2+(y-y_B)^2}),
  \end{equation*}
  where $v$ represents the speed of sound, and $(x_A,y_A)$ and $(x_B,y_B)$ are the locations of the observers.
  The bottom two restrictions in the diagram simply discard the thunder times, so
  \begin{equation*}
    \left(\shf{L}(A \to T)\right)(t_1,t_2) = t_1.
  \end{equation*}
  and
  \begin{equation*}
    \left(\shf{L}(B \to T)\right)(t_1,t_2) = t_1.
  \end{equation*}
  Unlike in the search and rescue sheaf (Example \ref{eg:titan_sheaf}), where commutativity was automatic (because all paths were of length $1$), now we need to check it,
  \begin{equation*}
    \begin{aligned}
      \left(\shf{L}(A \to T)\right) \circ \left(\shf{L}(X \to A)\right) (x,y,t) &= \left(\shf{L}(A \to T)\right) (t,t+1/v \sqrt{(x-x_A)^2+(y-y_A)^2}) \\
      &= t \\
      &= \left(\shf{L}(B \to T)\right) (t,t+1/v \sqrt{(x-x_B)^2+(y-y_B)^2}) \\
      &= \left(\shf{L}(B \to T)\right) \circ \left(\shf{L}(X \to B)\right) (x,y,t).\\
    \end{aligned}
  \end{equation*}
\end{example}

While the restrictions in the search and rescue and lightning examples are derived by physical assumptions,
this need not always be the case.
The binocular rivalry case instead encodes neurological connection assumptions.

\begin{example}(Overlapping vision sheaf)
  \label{eg:binocular_sheaf}
  We have two possible graphs that describe overlapping vision, from Examples \ref{eg:binocular_graph} and \ref{eg:binocular_topology},
  \begin{equation*}
    \xymatrix{
      &&    && V \ar[dr]\ar[dl]\ar[dd]\\
      L\ar[dr]&&R\ar[dl]    &L \ar[dr]  & & R \ar[dl]\\
      &B&    &&B&\\
      }
  \end{equation*}
  Since both agree on the lower portion of the graph, let us start there.
  
  Suppose for simplicity that each eye has $n$ rod receptors, and that $m$ rod receptors in each eye are able to see the overlap region between the two eyes.
  Each rod receptor senses a gray scale brightness value, which we will represent by a single real number in $\mathbb{R}$.
  The stalk $\shf{S}(L)$ should be interpreted as saying that the stalk over $L$ consists of a vector that lists the brightness sensed by each rod in the left eye.
  This suggests that a sheaf $\shf{S}$ describing the situation will have the following stalks
  \begin{equation*}
    \shf{S}(L) = \mathbb{R}^n, \; \shf{S}(R) = \mathbb{R}^n, \; \shf{S}(B) = \mathbb{R}^m.
  \end{equation*}
  
  The restrictions follow the pattern of the graph, so they fit into a diagram
  \begin{equation*}
    \xymatrix{
      \mathbb{R}^n \ar[dr] & & \mathbb{R}^n \ar[dl]\\
      & \mathbb{R}^m & \\
      }
  \end{equation*}
  The restrictions need to do two things:
  \begin{enumerate}
  \item Remove the values of receptors that are not able to sense the overlap region, and
  \item Perform a transformation to account for the fact that rod receptors in the left and right eyes that can sense the same part of the scene will receive different amounts of light due to geometry.
  \end{enumerate}
  These two requirements make the restrictions in $\shf{S}$ trace the neurological ``wiring'' of the optic nerves into the brain.
  While (1) consists of removing components in the vector, (2) is in general quite complicated (yet does not have a significant effect for images far from the eyes, so we will simply ignore this effect).
  For instance, if the stalk $\shf{S}(L)$ is organized so that the first $n-m$ values in $\mathbb{R}^n$ correspond to rod receptors that are unable to see the overlap,
  then the restriction map could simply be
  \begin{equation*}
    \left(\shf{S}(L \to B)\right) (\ell_1, \dotsc, \ell_{n-m}, \ell_{n-m+1} \dotsc \ell_n) = (\ell_{n-m+1}, \dotsc, \ell_n).
  \end{equation*}
  If we are using the smaller graph as our starting point, the above specification has completely defined our sheaf $\shf{S}$.
  As promised, the cospan $L \to B \leftarrow R$ corresponds to a test for consistency between the two eyes.
  The test is simply that the image projected on the retina from both eyes agrees on the overlap,
  \begin{equation*}
    \left(\shf{S}(L \to B)\right) (\ell_1, \dotsc, \ell_{n-m}, \ell_{n-m+1} \dotsc \ell_n) = (\ell_{n-m+1}, \dotsc, \ell_n) = (b_1, \dotsc, b_m),
  \end{equation*}
  and
  \begin{equation*}
    \left(\shf{S}(R \to B)\right) (r_1, \dotsc, r_{n-m}, r_{n-m+1} \dotsc r_n) = (r_{n-m+1}, \dotsc, r_n) = (b_1, \dotsc, b_m),
  \end{equation*}
  hence
  \begin{equation*}
    (\ell_{n-m+1}, \dotsc, \ell_n) = (b_1, \dotsc, b_m) = (r_{n-m+1}, \dotsc, r_n).
  \end{equation*}

  If we are instead using the larger graph that includes $V$, although the cospan $L \to B \leftarrow R$ still provides a test,
  there is more to the story, because we now get to define the span $L \leftarrow V \to R$.
  One simplistic definition (corresponding to the ``explanatory'' interpretation in the overlapping vision topology example (Example \ref{eg:binocular_topology})) is simply that
  \begin{equation*}
    \shf{S}(V) = \mathbb{R}^{2n - m},
  \end{equation*}
  which lists all of the rod receptors in both eyes, accounting for the ``duplicate'' receptors in the overlap.
  \end{example}

  But more interesting definitions are possible and worth considering, because spans build complex hypotheses.

  \begin{example} (Vision with depth)
    \label{eg:depth_sheaf}
  Taking the ``causal'' interpretation in the overlapping vision topology example (Example \ref{eg:binocular_topology}),
  we might suppose the visual field contains exactly one spherical object
  located at a position $\mathbb{R}^3$ (with whatever units and coordinate systems are convenient).
  Although context will usually assist the viewer in determining the size and distance to this spherical object,
  we will abstract that idea, and assume that the viewer knows the size of the object so that they can estimate its distance.
  We could therefore define
  \begin{equation*}
    \shf{S}(V) = \mathbb{R}^3.
  \end{equation*}
  In this case, the restrictions $\shf{S}(V \to L)$ and $\shf{S}(V \to R)$ do ``real work,''
  a more complex calculation than simply deleting items from a list.
  Because they are functions $\mathbb{R}^3 \to \mathbb{R}^n$, they predict the image projected on the retina from the spherical object, given its stated location.
  
  Implicit in the images on the retinas is the apparent size of the object.
  The restrictions $\shf{S}(L \to B)$ and $\shf{S}(R \to B)$ implicitly involve the estimated location of the object, derived from its apparent size and location on each retina individually.
  The restriction $\shf{S}(V \to B)$ provides a new consistency constraint on the model,
  because the these estimated from the left eye and the right eye must agree!
  This is expressed by the pair of equations
  \begin{equation*}
    \shf{S}(V \to B) = \shf{S}(L \to B) \circ \shf{S}(V \to L) = \shf{S}(R \to B) \circ \shf{S}(V \to R),
  \end{equation*}
  that enforce commutativity of the diagram.
  
  Suppose that $\shf{S}(B) = \mathbb{R}^m$ as before, representing the otherwise unprocessed scene in the overlap region.
  Parallax error will prevent the two eyes from seeing exactly the same view of the sphere,
  so it will be impossible for them to agree.
  If the sphere is far enough away, the parallax error will be small enough that it could be neglected.
  But in the strictest sense, \emph{choosing $\shf{S}(B) = \mathbb{R}^m$ breaks the sheaf commutativity condition}.

  A better choice is to define $\shf{S}(B) = \mathbb{R}^3$.
  The resulting sheaf diagram is
    \begin{equation*}
    \xymatrix{
      & \shf{S}(V) = \mathbb{R}^3 \ar[dr] \ar[dl] & \\
      \shf{S}(L) = \mathbb{R}^n \ar[dr] & & \shf{S}(R) = \mathbb{R}^n \ar[dl]\\
      & \shf{S}(B) = \mathbb{R}^3 & \\
      }
  \end{equation*}
  For this to work, the restrictions $\shf{S}(L \to B)$ and $\shf{S}(R\to B)$ imply that the optical system attempts to infer the position of the sphere from each eye \emph{independently} from its apparent size and position,
  and that these independent estimates are compared for consistency.
  More explicitly, the restriction $\shf{S}(L \to B)$, being a function $\mathbb{R}^n \to \mathbb{R}^3$,
  means that the left eye assesses not only the angular position of the sphere (its azimuth and elevation, for instance),
  but also its distance (from its apparent size).

  The interpretation of this final sheaf diagram is interesting, because although $\shf{S}(V)$ and $\shf{S}(B)$ are equal as sets,
  they mean different things.
  Specifically, being the center of a span,  $\shf{S}(V)$ represents the hidden latent model of external reality, which in this case consists of a single sphere.
  On the other hand, being the center of a cospan, $\shf{S}(B)$ represents a consistency test within the brain, as it tries to combine the scenes between the two eyes.
  We can then argue that the restriction $\shf{S}(V \to B)$ should be an identity function $\mathbb{R}^3 \to \mathbb{R}^3$, which says that the model of reality within the brain,
  constructed by each eye's depth perception, should agree with the external reality.
  This gives a strong notion of model self-consistency.
\end{example}

Failure of commutativity means that there is an ambiguity about how to make the prediction.

\begin{example}
  \label{eg:path_sheaf}
  Referring back to the road trip example (Example \ref{eg:path_planning}),
  attempting to build a sheaf means explaining exactly how the travel time predictions are computed.
  At each waypoint, we report the accumulated travel time in hours, which is a real number $\mathbb{R}$.
  When traveling from one waypoint to the next, we simply need to add the predicted travel time,
  which was written in Example \ref{eg:path_planning} by labeling the edge with a number of hours.
  Without belaboring the notation (especially because there would be many functions written out!),
  the proposed sheaf diagram (with restrictions summarized on the edges) is the following,
    \begin{equation*}
    \xymatrix{
      \mathbb{R} \ar[r]^{+3} & \mathbb{R} \ar[r]^{+1} &\mathbb{R}\ar[r]^{+1} & \mathbb{R} \ar[r]^{+1} & \mathbb{R} \ar[r]^{+1} \ar[d]_{+1} & \mathbb{R} \ar[d]^{+0.7} \ar[dl]_{+0.7} \\
                &            &           &            & \mathbb{R} \ar[dr]_{+1.1}\ar[r]^{+1} & \mathbb{R} \ar[d]^{+0.7} \\
                && \mathbb{R} \ar[dl]_{+2}\ar[d]^{+2} & \mathbb{R} \ar[l]_{+3} \ar[dl]^{+2}& \mathbb{R} \ar[d]^-{+0.8} \ar[l]_{+0.7}& \mathbb{R} \ar[l]^{+1.5}\ar@/^1pc/[dl]^{+1}\\
                & \mathbb{R}       & \mathbb{R}\ar[l]^{+1.7}          &                             & \mathbb{R} \ar[lu]^{+0.8} &\\
      }
    \end{equation*}

    This is \emph{not the diagram of a sheaf} because the commutativity requirement is not met.
    In particular, consider the subdiagram corresponding to $BOS$, $WOR$, and $PRD$
    \begin{equation*}
      \xymatrix{
        BOS \ar[r] \ar[dr] & WOR \ar[d] &\mathbb{R} \ar[r]^{+1} \ar[dr]_{+1.1} & \mathbb{R} \ar[d]^{+0.7} \\
        & PRD & &\mathbb{R}  \\
        }
    \end{equation*}
    The travel time estimate along the path $BOS \to WOR \to PRD$ is $1.7$ hours, but the estimate for traveling directly $BOS \to PRD$ is only $1.1$ hours.
    This means that the diagram does not commute, because starting at $BOS$ and ending at $PRD$ depends on the path taken.
\end{example}


\section{Consistency: Inferences from observations}
\label{sec:consistency}

In this section,
we will supply enough detail to a model so as to provide a quantitative test statistic for measuring the fit between a thus-far consistent model (a sheaf) and observations of the variables in the model.

\subsection{Perfect consistency: sections of a sheaf}

\begin{definition}
  \label{def:assignment}
  Suppose that $\shf{S}$ is a sheaf on a directed graph $G=(V,E)$,
  and that  $S \subseteq V$ is a collection of vertices.
  An \emph{assignment to $\shf{S}$ supported on $A$} is an element of the Cartesian product
  \begin{equation*}
    \prod_{x \in A} \shf{S}(x),
  \end{equation*}
  which can also be thought of as a vector of values indexed by the elements of $A$.
  If $A = V$, then we say the assignment is a \emph{global assignment}.
\end{definition}

The support of an assignment declares what was actually observed.
Vertices outside of the support are assumed to be latent, not observed directly, or otherwise inaccessible.
This does not have to be a hard and fast distinction.
We can start with a small support with actual observations, try to impute missing data, or try to infer latent variables,
and then use these iteratively to expand the support.

When an assignment is consistent on its support, it is called a \emph{section}.

\begin{definition}
  \label{def:section}
  Suppose that $G=(V,E)$ is a directed acyclic graph, and we endow $G$ with its partial order $\le$ (Corollary \ref{cor:poset}). 
  A \emph{section} of a sheaf $\shf{S}$ on a directed graph $G$ is an assignment $s$ supported on $A$ such that
  \begin{enumerate}
  \item $A$ is an open set in $Alex(G)$, which means that $A$ is upwardly closed, that is, if $x \in A$ and $x \le y$, then $y \in A$, and
  \item for all $x, y$ in $A$, such that $x \le y$, we have that $s_y$ can be computed from $s_x$ by applying the restriction $\shf{S}(x \to y)$, or equivalently,
    \begin{equation*}
      s_y = \left(\shf{S}(x \to y )\right)(s_x).
    \end{equation*}
  \end{enumerate}
\end{definition}

If there is a well defined way to predict variables, then there is a sense in which additional spans (not present in the graph but are present in the topology) can be predicted.

\begin{example} (Sections of both vision sheaves)
  \label{eg:binocular_section}
  Recall the overlapping vision sheaf (Example \ref{eg:binocular_sheaf}) and the vision with depth sheaf (Example \ref{eg:depth_sheaf}) to describe binocular rivalry.
  If we start with the smaller sheaf, with diagram
  \begin{equation*}
    \xymatrix{
      \shf{S}(L) = \mathbb{R}^n \ar[dr] & & \shf{S}(R) = \mathbb{R}^n \ar[dl]\\
      & \shf{S}(B) = \mathbb{R}^m & \\
      }
  \end{equation*}
  the set of assignments is
  \begin{equation*}
     \shf{S}(L) \times \shf{S}(R) \times \shf{S}(B) = \mathbb{R}^n \times \mathbb{R}^n \times \mathbb{R}^m \cong \mathbb{R}^{2n+m},
  \end{equation*}
  in which a typical element is a vector
  \begin{equation*}
    (\ell_1, \dotsc, \ell_n, r_1, \dotsc, r_n, b_1, \dotsc, b_m).
  \end{equation*}
  Not all of these assignments are consistent.
  The equations describing agreement between the left and right eye's views with the overlap view, namely
  \begin{equation*}
    \left(\shf{S}(L \to B)\right) (\ell_1, \dotsc, \ell_{n-m}, \ell_{n-m+1} \dotsc \ell_n) = (\ell_{n-m+1}, \dotsc, \ell_n) = (b_1, \dotsc, b_m),
  \end{equation*}
  \begin{equation*}
    \left(\shf{S}(R \to B)\right) (r_1, \dotsc, r_{n-m}, r_{n-m+1} \dotsc r_n) = (r_{n-m+1}, \dotsc, r_n) = (b_1, \dotsc, b_m),
  \end{equation*}
  are exactly the conditions that define a section.
  The set of all sections is a vector space of dimension $\mathbb{R}^{2n-m}$,
  since there is redundancy between the values of the receptors in the overlap.
  
  Notice something interesting: the set of sections we found is exactly the same as one of the possible choices for the stalk $\shf{S}(V)$ listed in the overlapping vision sheaf example (Example \ref{eg:binocular_sheaf}).
  \emph{The sections have recovered the stalk on the hidden span in the sheaf diagram!}

  We can now resolve the difference between the causal and explanatory interpretations highlighted in the overlapping vision example (Example \ref{eg:binocular_graph}).
  If we consider the sheaf with $V$ present, namely
  \begin{equation*}
    \xymatrix{
      & \shf{S}(V) = \mathbb{R}^{2n-m} \ar[dr] \ar[dl] \ar[dd] & \\
      \shf{S}(L) = \mathbb{R}^n \ar[dr] & & \shf{S}(R) = \mathbb{R}^n \ar[dl]\\
      & \shf{S}(B) = \mathbb{R}^m & \\
      }
  \end{equation*}
  this has the same set of sections as the previous sheaf.
  The explanatory and causal interpretations can be distinguished by the support of the assignment under discussion.
  The causal interpretation corresponds to the case where the scene $V$ is in the support of the assignment,
  and the explanatory interpretation corresponds to the case where $V$ is not in the support.
  If we suppose that the model and observations are consistent, this means that we take an assignment $a$ with its stated support and extend it into a section.

  Taking the causal interpretation, where $V$ is in the support, the value $a(V) \in \mathbb{R}^{2n-m}$ determines the values of the section at $L$, $R$, and $B$.
  In short, by extending this assignment to a section, the scene determines what the eyes see.
  
  Taking the explanatory interpretation, where $V$ is not in the support, both values $a(L) \in \mathbb{R}^n$ and $a(R) \in \mathbb{R}^n$ which agree on $B$ allow us to infer the value at $V$.
  Briefly, by extending this assignment to a section, the two compatible scenes from both eyes determine what the mind perceives.

  There is yet more modeling freedom.
  If we take the vision with depth sheaf (Example \ref{eg:depth_sheaf}),
  the one in which a single spherical object is in view and the eyes each independently assess distance to the sphere,
  the sheaf diagram we found was
  \begin{equation*}
    \xymatrix{
      & \shf{S}(V) = \mathbb{R}^3 \ar[dr] \ar[dl] \ar[dd] & \\
      \shf{S}(L) = \mathbb{R}^n \ar[dr] & & \shf{S}(R) = \mathbb{R}^n \ar[dl]\\
      & \shf{S}(B) = \mathbb{R}^3 & \\
      }
  \end{equation*}
  where we have not shown all of the restriction maps, because they can be inferred by commutativity.
  Since the assignments for this sheaf include the stalk $\shf{S}(V)$,
  they list the location of the sphere, the left eye's view, the right eye's view, and the inferred location of the sphere made independently by each eye.
  A section enforces consistency between all of these.
  Since $V$ sits at the top of the diagram in a span,
  each section is \emph{completely} determined by $\mathbb{R}^3$, the location of the sphere.
\end{example}

Above, examining the sections of both vision sheaves (Example \ref{eg:binocular_section}) showed that a candidate stalk for a hidden cospan can be inferred from the rest of the sheaf diagram.
This is a special case of a more general phenomenon, which (as in the example),
plays a useful role in distinguishing between causal and explanatory interpretations of a model.

Computing sections of sheaves is computationally rich enough to encode complex logical models.
In fact, nearly all of the examples in this chapter fall under the conditions discussed in \cite{Garey_1974} which establishes
that computing sections of sheaves is NP-complete.

\begin{example}
  \label{eg:logic_puzzle}
Consider a logic puzzle in which there are four characters, 
\begin{enumerate}
\item Alice (female),
\item Bob (male),
\item Chris (gender unclear), and
\item Dillon (gender unclear).
\end{enumerate}
Each of the characters is wearing a shirt of a different color, which could be Red, Yellow, Green, or White.

The logic puzzle specifies three clues,
\begin{enumerate}
\item The yellow shirt is on a male.

\item Chris and Dillon comment about the two others.  
    Chris says to Dillon:
    ``She'd look better in that green shirt he's wearing!''
    Dillon replies to Chris:
    ``I'm offended that you would make remarks about a woman's clothing. You guys don't understand us women!''

\item Alice comments that she loves the red shirt she is wearing.
\end{enumerate}

In order to encode this puzzle as a sheaf, we start by formalizing a proposed solution as a table with four rows (characters) and four columns (shirt colors).
One can fill each cell with a $1$ if the character is wearing the specified shirt color, or with a $0$ otherwise.
Since each character is wearing a shirt with one color, each row will sum to $1$.
Since each color shirt is on exactly one character, each column will sum to $1$ as well.

\begin{table}
  \begin{center}
    \caption{Table of values for the stalk over $M$ in Example \ref{eg:logic_puzzle}.  (Blank cells are not yet known given the clues.)}
    \label{tab:logic_puzzle}
    \begin{tabular}{|l|c|c|c|c|}
      \hline
      Character & Red & Yellow & Green & White\\
      \hline
      Alice     & 1   & 0      &    0  & 0 \\
      Bob       & 0   & 0      &    1  & 0 \\
      Chris     & 0   &        &    0  & \\
      Dillon    & 0   &        &    0  & \\
      \hline
    \end{tabular}
  \end{center}
\end{table}

The remaining clues are a little more subtle, and make use of the machinery we have built thus far.
Table \ref{tab:logic_puzzle} shows the table after noting information from clues (2) and (3);
blank entries are not yet known.
We start by building a graph with a single span, where the solution (unknown) is the central vertex,
  \begin{equation*}
    \xymatrix{
      R_0 &    &     & M \ar[lll]\ar[dlll]\ar[ddlll]\ar[dddlll] \ar[rr] \ar[drr]\ar[ddrr]\ar[dddrr]\ar[llddd]\ar[lddd]\ar[ddd]\ar[rddd]  &     & C_0\\
      R_1 &    &     &    &     & C_1\\
      R_2 &    &     &    &     & C_2\\
      R_3 &X_1 & X_2 & X_3 & X_4 & C_3\\
      }
  \end{equation*}
  The specification of the problem will start as a sheaf on the above graph \emph{and} an assignment supported on
  \begin{equation*}
    \{R_0, R_1, R_2, R_3, C_0, C_1, C_2, C_3, X_1, X_2, X_3, X_4\}.
  \end{equation*}
  The $R_n$ refer to sums of rows, the $C_n$ refer to sums of columns, and the $X_n$ refer to additional clues in the puzzle.
  The solution to the logic puzzle extends the assignment to a section supported on the entire graph.
  
  The stalks of our sheaf should be built from Boolean values $\{0,1\}$ or integers $\mathbb{Z}$.
  However, there is no harm in enlarging these sets to $\mathbb{R}$.
  Noting that there are $16$ entries in Table \ref{tab:logic_puzzle},
  we should define $\shf{S}(M) = \mathbb{R}^{16}$.
  To keep organized, we will read the entries in \emph{row major} order,
  starting with $0$.

  As an example, this means that the third entry in the vector of length $16$ refers to whether Alice is wearing Green,
  and the seventh entry refers to whether Bob is wearing Green.
  
  With this in mind, the sheaf has the following stalks,
  \begin{itemize}
  \item $\shf{S}(M) = \mathbb{R}^{16}$ for the solution,
  \item $\shf{S}(R_0) = \shf{S}(R_1) =\shf{S}(R_2) =\shf{S}(R_3) = \mathbb{R}$ for the row sums,
  \item $\shf{S}(C_0) = \shf{S}(C_1) =\shf{S}(C_2) =\shf{S}(C_3) = \mathbb{R}$ for the column sums, and
  \item $\shf{S}(X_1) = \shf{S}(X_2) =\shf{S}(X_3) =\shf{S}(X_4) = \mathbb{R}$ for the remaining conditions.
  \end{itemize}

  We can encode the row and column sums into the sheaf by using the condition for a section,
  since we can make an equation like,
  \begin{equation*}
    \begin{aligned}
      \left(\shf{S}(M \to R_0)\right)(m_0,\dots,m_{15}) &= s(R_0) \\
      \sum_{i=0}^3 m_{i} = m_0 + m_1+m_2+m_3 & = 1.\\
    \end{aligned}
  \end{equation*}
  Notice that this defines two pieces of data: the restriction map $\left(\shf{S}(M \to R_0)\right)$ and the value of the assignment at $R_0$.

  Repeating that structure, the row sums are enforced by a restriction map for $k=0, \dotsc, 3$,
  \begin{equation*}
    \left(\shf{S}(M \to R_k)\right)(m_0,\dots,m_{15}) = \sum_{i=0}^3 m_{4k+i},
  \end{equation*}
  with the constraint that the assignment takes the following values,
  \begin{equation*}
    s(R_0) = 1, \;     s(R_1) = 1, \;     s(R_2) = 1, \;    s(R_3) = 1. \;
  \end{equation*}
  Similarly, the column sums are enforced by the following restriction maps for $k=0, \dotsc, 3$,
  \begin{equation*}
    \left(\shf{S}(M \to C_k)\right)(m_0,\dots,m_{15}) = \sum_{i=0}^3 m_{k+4i},
  \end{equation*}
  along with the following constraints on the values of the assignment,
  \begin{equation*}
    s(C_0) = 1, \;     s(C_1) = 1, \;     s(C_2) = 1, \;    s(C_3) = 1. \;
  \end{equation*}
  
  Finally, we enforce a few more conditions by carefully chosen restriction maps and selecting values for part of the section we seek.
  The restriction map and assignment
  \begin{equation*}
    \left(\shf{S}(M \to X_1)\right)(m_0,\dots,m_{15}) = m_{6}, \; s(X_1) = 1,
  \end{equation*}
  follows from clue (2) stating that Bob wears Green.
  
  The pair of restriction maps and assignments
  \begin{equation*}
    \left(\shf{S}(M \to X_2)\right)(m_0,\dots,m_{15}) = m_{5} + m_{9}, \; s(X_2) = 1,
  \end{equation*}
  and
  \begin{equation*}
    \left(\shf{S}(M \to X_3)\right)(m_0,\dots,m_{15}) = m_5m_9, \; s(X_3) = 0,
  \end{equation*}
  together follow from clues (1) and (2), stating that the Yellow shirt is either on Bob or Chris (because clue (2) implies that Dillon is female).
  Finally, clue (3) is enforced by
  \begin{equation*}
    \left(\shf{S}(M \to X_4)\right)(m_0,\dots,m_{15}) = m_{0},
  \end{equation*}
  and
  \begin{equation*}
         s(X_4) = 1.
  \end{equation*}

  Given these restriction maps and assignment, the only value of an assignment that remains is on $M$.
  The only Boolean-valued vector that satisfies all of the equations specifying that this assignment is a section is
  \begin{equation*}
    s(M) = (1, 0, 0, 0,\;  0, 0, 1, 0,\;  0, 1, 0, 0,\;  0, 0, 0, 1),
  \end{equation*}
  This corresponds to the situation where
  \begin{itemize}
  \item Alice is wearing Red,
  \item Bob is wearing Green,
  \item Chris is wearing Yellow, and
  \item Dillon is wearing White.
  \end{itemize}

  Note that there are many other sections with non-Boolean values,
  but being a logic problem, only Boolean values make sense.
  If one attempts to write this sheaf in software using real values instead of Boolean ones,
  one might accidentally find one of these other sections.
  This is a sign that the puzzle is not properly constrained for non-Boolean values.
\end{example}

\begin{remark}
  It is possible to generalize from sheaves defined on graphs (as we do in this chapter) to sheaves defined on topological spaces.
  In particular, the definition of a sheaf on a directed acyclic graph defines a sheaf on its Alexandrov topological space (a fact that holds with substantial generality \cite{Curry_2013}).
  This is important when one needs to distinguish between causal and explanatory interpretations of a problem.
  Since topological spaces contain all unions of open sets, this means that there can be no hidden spans nor hidden cospans.
  Hence a sheaf built upon a topological space will necessarily support both interpretations, since it forces both spans and cospans to exist.
  On the other hand, if the goal is to build models and test them using statistical tools, this distinction plays a diminished role.
\end{remark}

\subsection{Handling approximation with pseudometrics}

The logic puzzle example (Example \ref{eg:logic_puzzle}) should give the reader confidence that sheaves and assignments are able to encode complex logical relationships,
and that the process of extending an assignment on a subgraph to a section is able to describe a wide variety of problems.
Nevertheless, Definition \ref{def:section} makes sections rather brittle: an assignment is or is not a section.
In many practical scenarios brittleness is undesirable, yet an approximate section is acceptable.
The traditional way to approximate data and transformations is through the use of \emph{continuity}.
If one can measure the size of a change, a function is \emph{continuous} if a small change on the input guarantees a small change on the output.
We measure the size of changes by the use of a \emph{pseudometric}.

\begin{definition}
  \label{def:continuity}
  If $X$ is a set, a \emph{pseudometric} $d$ is a function $d: X \times X \to [0,\infty)$ that satisfies the following properties
    \begin{description}
    \item[Reflexivity] $d(x,x) = 0$,
    \item[Symmetry] $d(x,y) = d(y,x)$,
    \item[Nonnegativity] $d(x,y) \ge 0$, and
    \item[Triangle inequality] $d(x,z) \le d(x,y) + d(y,z)$.
    \end{description}
    The pair $(X,d_X)$ is a \emph{pseudometric space}.
    
    If $d_X$ is a pseudometric on $X$ and $d_Y$ is a pseudometric on $Y$, and $f: X \to Y$ is a function,
    a function $f$ is \emph{Lipschitz continuous with constant $K$} if the distance between its outputs is less than $K$ times the distance between its inputs.
    In other words, 
    \begin{equation*}
      d_Y(f(x),f(x')) \le K \cdot d(x,x').
    \end{equation*}
\end{definition}

One can interpret continuity as a kind of geometric consistency check on the specification of a model,
because continuity ensures the stability of predictions made by the model.

\begin{example} (Vision model pseudometrics)
  \label{eg:binocular_continuity}
  In the case of the binocular rivalry examples, there are a few possible pseudometrics one might wish to consider.
  We have been treating the stalk $\shf{S}(L) = \mathbb{R}^n$ over the left eye as a vector, so it is natural to use the Euclidean distance.
  That is, if one considers two vectors of receptor values $(x_1, \dots, x_n)$, and $(x'_1, \dotsc, x'_n)$,
  the distance $d$ is the square root of the component-wise sum of squares, namely
  \begin{equation*}
    d((x_1, \dotsc, x_n), (x'_1, \dotsc, x'_n)) = \left(\sum_{i=1}^n |x_i - x'_i|^2\right)^{1/2}.
  \end{equation*}
  It is a classical argument (see \cite{kreyszig2011advanced}, for instance) that this is a pseudometric.
  
  In the overlapping vision sheaf (Example \ref{eg:binocular_sheaf}),
  there are three restrictions that should be investigated for continuity,
  \begin{equation*}
    \left(\shf{S}(L \to B)\right) (\ell_1, \dotsc, \ell_{n-m}, \ell_{n-m+1} \dotsc \ell_n) = (\ell_{n-m+1}, \dotsc, \ell_n),
  \end{equation*}
  \begin{equation*}
    \left(\shf{S}(R \to B)\right) (r_1, \dotsc, r_{n-m}, r_{n-m+1} \dotsc r_n) = (r_{n-m+1}, \dotsc, r_n),
  \end{equation*}
  and
  \begin{equation*}
    \shf{S}(V \to L) : \mathbb{R}^3 \to \mathbb{R}^n.
  \end{equation*}

  The first restriction (and its twin, naturally) is straightforward and can be handled directly,
  \begin{equation*}
    \begin{aligned}
      d\left(\left(\shf{S}(L \to B)\right)(\ell),\left(\shf{S}(L \to B)\right)(\ell')\right) & = d\left((\ell_{n-m+1}, \dotsc, \ell_n),(\ell'_{n-m+1}, \dotsc, \ell'_n)\right)\\
      & = \left( \sum_{i=1}^m |\ell_{n-m+i} - \ell'_{n-m+i}|^2\right)^{1/2}\\
      & \le \left( \sum_{i=1}^n |\ell_{i} - \ell'_{i}|^2\right)^{1/2}\\
      & \le d\left(\ell,\ell'\right),
    \end{aligned}
  \end{equation*}
  which reveals that $\shf{S}(L \to B)$ is Lipschitz with constant $1$.

  If we consider the vision with depth sheaf (Example \ref{eg:depth_sheaf}), involving the spherical object,
  then the restriction $\shf{S}(V \to L)$ (and its twin $\shf{S}(V \to R)$) is more subtle.
  Evidently, the Euclidean distance makes sense for $\mathbb{R}^3$, the location of the sphere in the scene, since it is an physical distance.
  The definition of Lipschitz continuity for this restriction,
  \begin{equation*}
    \begin{aligned}
      d\left(\left(\shf{S}(V \to L)\right)(x,y,z),\left(\shf{S}(V \to L)\right)(x',y',z')\right) \le& K \left(|x-x'|^2 + |y-y'|^2 \right.\\&\left.+ |z-z'|^2\right)^{1/2} \\
      =& Kr,\\
    \end{aligned}
  \end{equation*}
  means that if we move the sphere by $r$ units, the vector of brightness values received by the left eye cannot change by more than $Kr$ units.
  For small movements $r$ of the sphere, the perceived scene can only change by small amounts.
  Although we have not defined an equation for $\shf{S}(V \to L)$ or $\shf{S}(L \to B)$ in this chapter\footnote{We certainly could make such a definition, but the complexity of any realistic definition would obscure the point we are making here.}, any reasonable definition will have this property.
\end{example}

\subsection{Relaxing consistency under uncertainty}

With these preliminaries about continuity established,
we can finally define the main test statistic for a model specified by a sheaf, the \emph{consistency radius}.
The consistency radius relaxes the equation in Definition \ref{def:section} for a section
from strict equality to a numerical value computed using pseudometrics.
When one has a section, and Definition \ref{def:section} is therefore satisfied, the consistency radius is $0$.

\begin{definition}
  \label{def:consistency_radius}
  A sheaf $\shf{S}$ on a directed acyclic graph $G=(V,E)$ in which each stalk is a pseudometric space and each restriction is Lipschitz continuous with constant $K$,
  will be called a \emph{sheaf of pseudometric spaces}.

  If $a$ is an assignment to a sheaf $\shf{S}$ of pseudometric spaces on $G$ supported on an open set $U$ in $Alex(G)$,
  the \emph{$k$-norm consistency radius of $a$} is the quantity given by
  \begin{equation*}
    c_{\shf{S}}(a,U) := \left(\sum_{x \in U,\;} \sum_{y: x \le y} \left(w_y d_y(\shf{S}(x \le y)a_x,a_y)\right)^k \right)^{1/k},
  \end{equation*}
  where $w_y$ is a positive real number called the \emph{weight of $y$}.
  For $k=\infty$, because $\infty$-norms are defined using maxima, we define
  \begin{equation*}
    c_{\shf{S}}(a,U) := \max_{x \in U,\;} \max_{y: x \le y} w_y d_y(\shf{S}(x \le y)a_x,a_y).
  \end{equation*}
  
  Note that the fact that $U$ is open ensures that the consistency radius is well-defined.
  When emphasis is needed, we say $c_{\shf{S}}(a,U)$ is a \emph{local consistency radius} if $U \subset V$ and a \emph{global consistency radius} if $U = V$.
\end{definition}

Consistency radius is a consistency check on an assignment against the model provided by the sheaf.
Where the underlying graph has spans, a small consistency radius ensures that the latent model functionally determines the observations.
When the graph has cospans, the consistency radius gives rise to general hypothesis tests.

We will typically use the usual geometric distance ($k=2$) and set the weights to specific values determined by the problem (to normalize variances, for instance, see Example \ref{eg:chi2}),
since then the sheaf framework we are building will generalize familiar concepts like variance and standard deviation.
Minimizing consistency radius over a sheaf with a span is a general way to perform statistical inference.

\begin{example} (Least squares regression, reprise)
  \label{eg:linear_regression}
  Recall the linear regression example from Section \ref{sec:linear_example},
  in which we wanted to find the best fit line $L$ for four points, $A$, $B$, $C$, $D$,
  where
  \begin{equation*}
    A = (0,y_A),
  \end{equation*}
  \begin{equation*}
    B = (1.5,y_B),
  \end{equation*}
  \begin{equation*}
    C = (1.5,y_C),
  \end{equation*}
  and
  \begin{equation*}
    D = (3,y_D).
  \end{equation*}
  We previously found that the sheaf diagram for this problem is
  \begin{equation*}
    \xymatrix{
      && \mathbb{R} &&\\
      \mathbb{R} && \mathbb{R}^2 \ar[u]^{b} \ar[ll]_{1.5m+b} \ar[rr]^{1.5m+b} \ar[d]_{3m+b} && \mathbb{R} \\
      && \mathbb{R} &&\\
      }
  \end{equation*}
  and the regression problem consists of finding the best way to extend the assignment $a$
  \begin{equation*}
    a(A) = y_A, \; a(B) = y_B, \; a(C) = y_C, \; a(D) = y_D.
  \end{equation*}
  to take a value at $L$.

  By the ``best way to extend,'' we mean the assignment should minimize the sum of squares of the errors at each data point.
  We now can show that this is actually just the $2$-norm consistency radius with all $1$ weights,
  as was claimed in Section \ref{sec:linear_example}.
  We can compute this quantity directly,
  \begin{equation*}
    \begin{aligned}
      c_{\shf{L}}(a) =& \left(\sum_{x \in U,\;} \sum_{y: x \le y} \left(d_y(\shf{S}(x \le y)a_x,a_y)\right)^2 \right)^{1/2}\\
      =& \left(\left(d_A(\shf{S}(L \le A)(m,b),y_A)\right)^2 + \left(d_B(\shf{S}(L \le B)(m,b),y_B)\right)^2 +\right. \\ & \left. \left(d_C(\shf{S}(L \le C)(m,b),y_C)\right)^2 + \left(d_D(\shf{S}(L \le D)(m,b),y_D)\right)^2\right)^{1/2}\\
      =& \left(\left(d_A(b,y_A)\right)^2 + \left(d_B(1.5m+b,y_B)\right)^2 + \left(d_C(1.5m+b,y_C)\right)^2 + \left(d_D(3m+b,y_D)\right)^2\right)^{1/2}\\
      =& \left(|b-y_A|^2 + |1.5m+b-y_B|^2 + |1.5m+b-y_C|^2 + |3m+b-y_D|^2\right)^{1/2},\\
    \end{aligned}
  \end{equation*}
  which is exactly the classical sum of squares of the errors.
\end{example}

With this framing,
the binocular rivalry problem is now a more elaborate framing of a missing data inference problem.

\begin{example}(Overlapping vision consistency radius)
  \label{eg:binocular_consistency}
  The vision model pseudometric example (Example \ref{eg:binocular_continuity}) already advocated choosing the Euclidean distance for the pseudometric in the binocular rivalry sheaf examples,
  so let us continue this.
  Start with the smaller sheaf, with diagram
  \begin{equation*}
    \xymatrix{
      \shf{S}(L) = \mathbb{R}^n \ar[dr] & & \shf{S}(R) = \mathbb{R}^n \ar[dl]\\
      & \shf{S}(B) = \mathbb{R}^m & \\
      }
  \end{equation*}
  and an assignment $a$ supported on $\{L,R,B\}$ given by
  \begin{equation*}
    a= (\ell_1, \dotsc, \ell_n, r_1, \dotsc, r_n, b_1, \dotsc, b_m).
  \end{equation*}
  Let us choose all of the weights to be $1$ for simplicity.
  The $k$-norm consistency radius for this assignment can be computed directly by unpacking Definition \ref{def:consistency_radius}
  step-by-step,
  \begin{equation*}
    \begin{aligned}
      c_{\shf{S}}(a,\{L,R,B\}) =& \left(d_{\{B\}}\left(\left(\shf{S}(L\to B)\right)(\ell_1,\dotsc,\ell_n),(b_1,\dotsc,b_m)\right)^k\right. \\&\left.+ d_{\{B\}}\left(\left(\shf{S}(R \to B)\right)(r_1,\dotsc,r_n),(b_1,\dotsc,b_m)\right)^k\right)^{1/k}\\
      =& \left( \left(\sum_{i=1}^m |\ell_{n-m+i}-b_i|^2 \right)^{k/2} + \left(\sum_{i=1}^m |r_{n-m+i}-b_i|^2\right)^{k/2} \right)^{1/k}\\
    \end{aligned}
  \end{equation*}
  This is simply measuring the difference between the scenes from the left eye and right eye in the overlap region.
  If there is no difference at all (a section), then all of the terms above are zero.

  It is also enlightening to consider the case where the assignment is supported on $\{L\}$, which models the situation where the right eye is covered.
  (In the Alexandrov topology, $\{L\}$ is not an open set, since any open set containing $L$ must also contain $B$.)
  The assignment is expressed by the vector
  \begin{equation*}
    a = (\ell_1, \dotsc, \ell_n).
  \end{equation*}
  Definition \ref{def:consistency_radius} for the consistency radius does not directly apply to this situation because $\{L\}$ on its own is not an open set.
  The smallest open set containing $\{L\}$ is $\{L,B\}$, so we can ask for the assignment on $\{L,B\}$ that ``best'' extends $a$ to an assignment on $b$.
  This ``best'' choice is an assignment with minimal consistency radius.
  We can reuse the calculation from above to reason that there is only one set of terms that correspond to the set $\{L,B\}$,
  and that all of the other terms do not apply.  Removing those irrelevant terms leaves,
  \begin{equation*}
    c_{\shf{S}}(a,\{L,B\}) = \left( \left(\sum_{i=1}^m |\ell_{n-m+i}-b_i|^2 \right)^{k/2} \right)^{1/k}.
  \end{equation*}
  Notice that this is minimized precisely when we define
  \begin{equation*}
    b_i = \ell_{n-m+i},
  \end{equation*}
  by simply re-using the receptor values from the left eye \emph{as} the values in the overlap.
  In fact, this is a section supported on $\{L,B\}$.
  Moreover, we can ask for extensions of this new assignment to $\{L,R,B\}$.
  Instead of completely determining \emph{all} of the values of $(r_1,\dotsc, r_n) \in \shf{S}(R) = \mathbb{R}^n$,
  only the last $m$ values of such a vector will be determined.
  This means that the perceived image in the right eye that is most consistent with the model we have built partly contains a copy of the image from the left eye.
  Effectively, the consistency radius-minimizing assignments model the brain synthesizing a fictitious image in the right eye's view.

  In a similar way, what if we were to have started with an assignment on $\{L,R\}$ instead of $\{L\}$?
  Although a little more tricky mathematically, this is perhaps more realistic than starting by assuming a value on $B$.
  Again, the set $\{L,R\}$ is not open, but we can ask for the assignment with smallest consistency radius that extends into this open set.
  In essence, the problem is now a missing data problem, wherein we seek to find a way to extend the assignment to an assignment on $B$ that yields the smallest consistency radius.
  Namely, we will have to find the vector $(b_1, \dotsc, b_n)$ (just the values in the overlap region) that solves\footnote{The $\argmin$ function finds the arguments which minimize the specified expression.}
  \begin{equation*}
    \argmin_{(b_1, \dotsc, b_n)} c_{\shf{S}}(a,\{L,R,B\}).
  \end{equation*}
  If $k=2$, this problem has a standard solution, 
  \begin{equation*}
    b_i = \frac{\ell_{n-m+i} + r_{n-m+i}}{2},
  \end{equation*}
  which makes the consistency radius half the Euclidean distance between the two eye's vectors, when restricted to the overlap.
\end{example}


\begin{remark}
  The term ``consistency radius'' is best read as ``the radius of a ball centered on the assignment that cannot contain a section.''
  This is true (up to Lipschitz constants), and is established in \cite{Robinson_sheafcanon}.
\end{remark}

\section{Testing consistency as a hypothesis}
\label{sec:testing}

We now are in a position to handle hypotheses concerning distributions.
There are two possible strategies: to make a sheaf of distributions, or to put a distribution on assignments.
We take the latter approach, since it is closer to what can be estimated from observational data.
With this perspective, consistency radius is a test statistic for the data as a sample (assignment) from a null model (sheaf and assignment distribution).

\begin{definition}
  \label{def:p_value}
  An \emph{assignment distribution} for a sheaf $\shf{S}$ on $(X,\col{T})$ is a probability measure on its space of assignments, $\prod_{U \in \col{T}} \shf{S}(U)$.
  Given a sheaf $\shf{S}$ of pseudometric spaces on $(X,\col{T})$ and an assignment distribution $\mu$,
  the \emph{consistency distribution} is the probability of an assignment yielding a given consistency radius when the assignments are drawn from $\mu$.
  Concretely, the consistency distribution states that the probability that the consistency radius is within an interval $(r_1,r_2) \subseteq [0,\infty)$,
    is given by
    \begin{equation*}
      \mu \left\{ a \in \prod_{U \in \col{T}} \shf{S}(U) : r_1 \le c_{\shf{S}}(a,X) \le r_2 \right\}.
    \end{equation*}

    If $a$ is an assignment to $\shf{S}$,
    its \emph{$p$-value} is the probability that the consistency radius would be at least as large as $c_{\shf{S}}(a,X)$,
    under the model given by the assignment distribution $\mu$ on the sheaf $\shf{S}$.
\end{definition}

The consistency distribution is stable, which ensures its usefulness as a tool.

\begin{proposition}
  The consistency distribution is weakly continuously dependent upon the assignment distribution.
\end{proposition}

\begin{proof}
  Category theorists will notice that the consistency distribution is the pushforward of $\mu$ along the consistency radius function $c: \prod_{U \in \col{T}} \shf{S}(U) \to [0,\infty)$.
  Lipschitz continuity of the consistency radius function is established in \cite{robinson2020assignments}.
  The ``Portmanteau theorem'' establishes that weak continuity for measures is preserved by pushforwards along a Lipschitz continuous function (standard, see \cite{Kenke_2006} for example).
\end{proof}

As a warm-up to handle the binocular rivalry case, let us show that classical $\chi^2$ test for the mean arises from a particular sheaf and assignment distribution.

\begin{example} (Goodness of fit test with consistency radius)
  \label{eg:chi2}
  Consider a single cospan of the form
  \begin{equation*}
    \xymatrix{
      A_1 \ar[dr] & \dotsb & A_n \ar[dl]\\
      & M & \\
      }
  \end{equation*}
  
  The interpretation will be that the $A_i$ represent observations in a sample, and $M$ represents the mean $\mu$ of the true distribution.
  We will also assume that we know the standard deviation $\sigma$ of the true distribution.
  We will use the sheaf diagram
  \begin{equation*}
    \xymatrix{
      \mathbb{R} \ar[dr] & \dotsb & \mathbb{R} \ar[dl]\\
      & \mathbb{R} & \\
      }
  \end{equation*}
  in which the arrows represent restrictions that are identity functions, $f(x) = x$.

  We now need to define the assignment distribution.
  Write the set of assignments
  \begin{equation*}
    (a_1, \dotsc, a_n, m).
  \end{equation*}
  Evidently, each of the $a_i$ are independently distributed with mean $\mu$ and standard deviation $\sigma$.
  Now since we propose that the mean of the distribution is $\mu$,
  we should place all probability mass on $m=\mu$, so that any assignment for which $m \not= \mu$ will have probability zero.
  
  If we choose the weight to be $1/\sigma$ in the $k=2$-norm consistency radius, this results in a consistency radius
  \begin{equation*}
    c\left((a_1,\dotsc, a_n,m=\mu),\{A_1,\dotsc,A_n,M\}\right)) = \left(\sum_{i=1}^n \left(\frac{|a_i - \mu|}{\sigma}\right)^2 \right)^{1/2}.
  \end{equation*}
  The square of this value is distributed according to the classical $\chi^2$ distribution with $n$ degrees of freedom.

  Moreover, if $\mu$ is not known, this corresponds to the situation where the assignment is supported only on $\{A_1, \dotsc, A_n\}$.
  The minimizer of consistency radius for such an assignment recovers the sample mean
  \begin{equation*}
    m = \frac{1}{n}\sum_{i=1}^n a_i.
  \end{equation*}

  From this it is immediate that since the consistency radius is $\chi^2$ distributed,
  the resulting $p$-value is exactly the chance under the hypothesis built by this sheaf that the sample would have a consistency radius that large or larger.
  This is exactly the classical $\chi^2$ goodness-of-fit test.
\end{example}

Consistent with the theme throughout this chapter, since cospans test hypotheses, the $\chi^2$ test appears as a single cospan.
Now with the classical $\chi^2$ test in hand, we return to the binocular rivalry case one final time.

\begin{example} (Overlapping vision $p$-values)
  \label{eg:binocular_pvalue}
  Recall the binocular rivalry examples from earlier in the chapter.
  Consider the overlapping vision sheaf (Example \ref{eg:binocular_sheaf}), namely
  \begin{equation*}
    \xymatrix{
      \shf{S}(L) = \mathbb{R}^n \ar[dr] & & \shf{S}(R) = \mathbb{R}^n \ar[dl]\\
      & \shf{S}(B) = \mathbb{R}^m & \\
      }
  \end{equation*}
  As noted many times in this chapter, the cospan yields the test for the hypothesis posed by the span,
  that the two eyes' images agree on the overlap to form a unified image.
  To use Definition \ref{def:p_value} to run this test,
  we need an assignment distribution.
  The set of assignments for this sheaf is given by
  \begin{equation*}
    \shf{S}(L) \times \shf{S}(R) \times \shf{B} = \mathbb{R}^n \times\mathbb{R}^n \times\mathbb{R}^m,
  \end{equation*}
  with elements
  \begin{equation*}
    (\ell_1, \dotsc, \ell_n, r_1, \dotsc, r_n, b_1, \dotsc, b_m).
  \end{equation*}
  In principle, the assignment distribution can be any probability distribution over this set.
  But for most of these distributions, the resulting distribution of consistency radius would be uninformative.

  Instead, a much more informative choice is to consider a multivariate distributed Gaussian distribution centered on the subspace of sections.
  One can interpret this as meaning that we expect the typical assignment to be close to a section, where both eyes are seeing similar images.
  For purposes of discussion, let us suppose that each receptor is independent of the others,
  and that the variances of the distribution depend on the stalk only, and are block-wise identically distributed.
  To that end, suppose that the standard deviation of this noise is given by $\sigma_L$, $\sigma_R$, and $\sigma_B$, respectively.
  
  Unlike the goodness of fit test with consistency radius (Example \ref{eg:chi2}), where there was a fixed mean that applied to every sample,
  there is a dependency structure within the assignment distribution in this case.
  That dependency is somewhat benign, and as a consequence,
  \begin{equation*}
    d_B(\shf{S}(L \to B)a(L),a(B))^2 = \sum_{i=1}^m |\ell_{n-m+i}-b_i|^2
  \end{equation*}
  is a sum of squares of Gaussians with variance $(\sigma_L^2 + \sigma_B^2)$,
  which is sometimes known as a generalized $\chi^2$ distribution with $m$ degrees of freedom.
  When we write the square of the $k=2$-norm consistency distribution using this variance in the weight,
  \begin{equation*}
    c_{\shf{S}}(a,\{L,R,B\})^2 = \sum_{i=1}^m \left(\frac{|\ell_{n-m+i}-b_i|^2}{\sigma_L^2+\sigma_B^2} +\frac{|r_{n-m+i}-b_i|^2}{\sigma_R^2+\sigma_B^2}\right),
  \end{equation*}
  this is $\chi^2$ distributed with $2m$ degrees of freedom.
\end{example}


\section{Future directions}

\subsection{Multiple sub-hypotheses}

Sheaves with assignments extend classical multiple hypothesis testing,
because there is a consistency radius (and hence a hypothesis test) for each open set.

\begin{definition}
  Given a significance level $\alpha$, \emph{family-wise error rate (FWER)} is the probability that at least one local consistency radius exceeds $\alpha$.
\end{definition}

Classical control of FWER for multiple hypotheses can once again be modeled by a sheaf on a cospan.

\begin{example}
  \label{eg:bonferroni}
  Consider the same sheaf $\shf{S}$ as for the goodness of fit test with consistency radius example (Example \ref{eg:chi2}) on the following graph
  \begin{equation*}
    \xymatrix{
      A_1 \ar[dr] & \dotsb & A_n \ar[dl]\\
      & M & \\
      }
  \end{equation*}
  in which the stalks are
  \begin{equation*}
    \shf{S}(M) = \shf{S}(A_1) = \dotsb = \shf{S}(A_n) = \mathbb{R}.
  \end{equation*}
  For simplicity, suppose each of the restriction maps are identity functions,
  \begin{equation*}
    \shf{S}(A_k \to M) (x) = x.
  \end{equation*}
  There are many open sets in the Alexandrov topology for this sheaf, but all of them are unions of sets of the form
  \begin{equation*}
    \{M\}, \{M,A_1\}, \dotsc, \{M,A_n\}.
  \end{equation*}
  Each one of these ultimately will correspond to a consistency radius, and hence a hypothesis test.
  
  Choose an assignment distribution in which the stalk distributions on $A_1$ through $A_n$ are all independent and identically distributed,
  and has a point mass on $M$, which is to say that the
  probability distribution function is
  \begin{equation*}
    p(a_1, \dotsc, a_n, b ) = \begin{cases}
      p(a_1) \dotsb p(a_n) & \text{if }b=0\\
      0 & \text{otherwise.}\\
    \end{cases}
  \end{equation*}

  If we have an assignment $(a_1, \dotsc, a_n, b)$, its consistency radius on the open set $\{M, A_k\}$ is
  \begin{equation*}
    c_{\shf{S}}((a_1, \dotsc, a_n, b), \{M,A_k\}) = d_M(\shf{S}(A_k \to B)(a_1, \dotsc, a_n, b), b) = |a_k - b|.
  \end{equation*}
  However, notice that given the assignment distribution we chose, the probability that $b$ takes a value other than $0$ is $0$.
  Therefore, the consistency distribution for this open set has probability distribution function $p(|a_k|)$.
  
  Suppose now we select a significance threshold $\alpha$.
  The FWER is the probability that there is at least one open set $\{M,A_K\}$ for which the consistency radius exceeds $\alpha$.
  \begin{equation*}
    \begin{aligned}
      FWER =& 1 - \left(1 - p(|a_1| > \alpha)) \dotsb (1-p(|a_n| < \alpha)\right) \\
      \approx& \left(p(|a_1| > \alpha) + \dotsb + p(|a_n| > \alpha)\right)\\
      \approx& n p(|a_1| > \alpha).\\
    \end{aligned}
  \end{equation*}
  On the other hand, we can also write
  \begin{equation*}
    FWER = p (\max_{k=1}^n |a_k| > \alpha).
  \end{equation*}
  Since the $\infty$-norm consistency radius is
  \begin{equation*}
    c_{\shf{S}}((a_1, \dotsc, a_n, b), \{M,A_1,\dotsc,A_n\}) = \max_{k=1}^n |a_k|,
  \end{equation*}
  this means that the global consistency radius exceeds $\alpha$ whenever any $|a_k| > \alpha$.
\end{example}

Notice that
\begin{equation*}
  FWER \approx \left(p(|a_1| > \alpha) + \dotsb + p(|a_n| > \alpha)\right)
\end{equation*}
is the classical \emph{Bonferroni approximation} used in multiple hypothesis testing.
Example \ref{eg:bonferroni} establishes that \emph{consistency radius generalizes the Bonferroni approximation} for multiple hypotheses.
Moreover,
\begin{enumerate}
\item The local consistency radii give each test in the multiple hypothesis testing framework, and
\item The $\infty$-norm global consistency distribution measures the FWER.
\end{enumerate}

If the sheaf in Example \ref{eg:bonferroni} has general linear restriction maps,
the assignment distribution is independent across stalks,
but now the stalk on the central node has a general distribution,
then we recover Bonferroni's approximation for FWER only when we condition upon the central node's stalk.
This is an expression of Judea Pearl's ``collider.''

Moreover, the diagrammatic nature of sheaves allows you to design more subtle dependencies into the null model.
If we consider a sheaf on
\begin{equation*}
  \xymatrix{
    A_n \ar[r] &\dotsb \ar[r] & A_3 \ar[r] & A_2 \ar[r] & A_1
  }
\end{equation*}
with linear restriction maps and independent stalks,
the consistency radii on successively larger open sets $\{A_1\}$, $\{A_1,A_2\}$, $\dotsc$ are monotonically increasing.
Therefore, even though FWER is not well approximated by a sum of probabilities, because there is dependence between the consistency distributions,
the global consistency radius still measures the FWER.
In short, the set of tests against consistency radius admits considerable flexibility,
much of it yet unexplored.

\subsection{Building informative null models}

Consistency radius measures the amount of disagreement between model and observations,
while the \emph{consistency filtration} \cite{robinson2020assignments},
which lists the open sets on which consistency radius is below a specified amount,
localizes the disagreement to specific open sets.

The consistency filtration appears to play the role of a generalized confidence interval.  
When you pick a significance level,
this sets a consistency filtration level,
and thereby selects where the the sheaf and assignment are or are not fitting.
The consistency filtration presents a chain of acceptance regions for a single inferential problem (the entire sheaf and the assignment to it).

Of course, if you want tests to be informative, the null distribution needs to have enough entropy.
This in principle exists.

\begin{proposition}
  Given a sheaf of finite pseudometric spaces on a finite partially ordered set with continuous restriction maps,
  there is a distribution on assignments
  such that the distribution of global consistency radius has maximum entropy.
\end{proposition}
We might be able to lift one or more of the finiteness assumptions.
Moreover, the distance between the maximum entropy distribution and the actual assignment distribution is a measure of how easy the model is to test.
\begin{proof}
  Since the base space of the sheaf is finite,
  and the stalks are also finite,
  the space of probability distributions on assignments is
  given by a closed finite dimensional simplex.
  This is hence also compact,
  because the simplex is bounded.
  Since the restriction maps are continuous,
  so is the consistency radius \cite[Thm. 1]{robinson2020assignments}.
  The space of distributions of consistency radius is the pushforward of the assignment distribution
  along the consistency radius function.
  The entropy of the distribution is furthermore a continuous function of this pushforward distribution.
  Therefore, the function taking a probability distribution on assignments to its entropy is continuous,
  hence it attains a maximum on its (compact) domain.
\end{proof}

\section{Conclusion}

In this chapter, we have built a tower of increasing consistency constraints for models,
summarized in Table \ref{tab:outline}.
Each new assumption becomes a fulcrum for additional inferential leverage:
\begin{itemize}
\item    Directed graphs yield the ability to exclude impossible dependencies,
\item    Topology identifies admissible comparisons,
\item    Sheaves enforce compatibility between state variables,
\item    Continuity ensures stability of predictions,
\item    Consistency radius yields a quantitative goodness of fit, and
\item    The consistency distribution provides rigorous hypothesis testing.
\end{itemize}
We have demonstrated how these work on a variety of examples,
carrying forth the idea that models require a notion of internal consistency,
and that they can be tested against observational data systematically.

\section*{Acknowledgments}

This is a preprint of the following chapter: Dietz, Robinson, ``Statistical Sheaf Modeling Methods for Measuring Qualia Structures'',
to published in \emph{Qualia Structure},
edited by Naotsugu Tsuchiya, Masafumi Oizumi, Yusuke Moriguchi, and Makiko Yamada, published by Springer Nature.
It is the version of the author's manuscript prior to acceptance for publication and has not
undergone editorial and/or peer review on behalf of the Publisher (where applicable).

This article is based upon work supported by the Office of Naval Research (ONR) under Contract No. N00014-25-1-2546.
Any opinions, findings and conclusions or recommendations expressed in this article are those of the authors and do not necessarily reflect the views of ONR.

\section*{AI Disclosure}
Microsoft Copilot was consulted for editorial and structural suggestions during the writing of this chapter,
and for suggesting standard references for facts already well known to the authors.
The text and mathematical content was written entirely by the authors.
No AI was used in producing mathematical content.

\bibliographystyle{plain}
\bibliography{qualiabook_bib}

\end{document}